\documentclass[useAMS,usenatbib]{mn2e}
\usepackage{graphicx}
\usepackage{times}
\usepackage{amssymb}
\usepackage{natbib}
\usepackage{pgf}
\usepackage{bm}
\def\gsim { \lower .75ex \hbox{$\sim$} \llap{\raise .27ex \hbox{$>$}} }
\def\lsim { \lower .75ex \hbox{$\sim$} \llap{\raise .27ex \hbox{$<$}} }
\begin{document}

\title[Feedback and the structure of galaxies at z=2]
{Feedback and the Structure of Simulated Galaxies at redshift z=2}

\author[Sales et al.]{
\parbox[t]{\textwidth}{
Laura V. Sales$^{1}$, 
Julio F. Navarro$^{2,3}$\thanks{Fellow of the Canadian Institute for Advanced Research.}, 
Joop Schaye$^{4}$, 
Claudio Dalla Vecchia$^{4,5}$,
Volker Springel$^{6,7,8}$ 
and C. M. Booth$^{4}$
}
\\
\\
$^{1}$ Kapteyn Astronomical Institute, P.O. Box 800, Groningen, The Netherlands\\
%\\
$^{2}$ Department of Physics and Astronomy, University of Victoria, Victoria, BC V8P 5C2,
Canada\\
$^{3}$ Department of Astronomy, University of Massachusetts, US\\
$^{4}$ Leiden Observatory, Leiden University, PO Box 9513, 2300 RA Leiden, The Netherlands\\
%\\
$^{5}$ Max-Planck-Institute for Extraterrestrische Physik, Giessenbachstrasse, D-85478, Garching, Germany\\
$^{6}$Max Planck Institute for Astrophysics, Karl-Schwarzschild-Strasse 1, 85740 Garching, Germany\\
$^{7}$Heidelberg Institute for Theoretical Studies,
  Schloss-Wolfsbrunnenweg 35, 69118 Heidelberg, Germany\\
$^{8}$Zentrum f\"{u}r Astronomie der Universit\"{a} Heidelberg,
    M\"onchhofstr. 12–14, 69120 Heidelberg, Germany\\
}

\maketitle

\begin{abstract} We study the properties of simulated high-redshift
  galaxies using cosmological N-body/gasdynamical runs from the
  OverWhelmingly Large Simulations ({\small OWLS}) project. The runs
  contrast several feedback implementations of varying effectiveness:
  from no-feedback, to supernova-driven winds to
  powerful AGN-driven outflows.  These different feedback models
  result in large variations in the abundance and structural
  properties of bright galaxies at $z=2$. In agreement with earlier
  work, models with inefficient or no feedback lead to the formation
  of massive compact galaxies collecting a large fraction (upwards of
  $50\%$) of all available baryons in each halo. Increasing the
  efficiency of feedback reduces the baryonic mass and increases the
  size of simulated galaxies. A model that includes 
  supernova-driven gas outflows aided by the energetic output of AGNs
  reduces galaxy masses by roughly a factor of $\sim 10$ compared with
  the no-feedback case. Other models give results that straddle these
  two extremes. Despite the large differences in galaxy formation
  efficiency, the net specific angular momentum of a galaxy is, on
  average, roughly half that of its surrounding halo, independent of
  halo mass (in the range probed) and of the feedback scheme. Feedback
  thus affects the baryonic mass of a galaxy much more severely than
  its spin. Feedback induces strong correlations between angular
  momentum content and galaxy mass that leave their imprint on galaxy
  scaling relations and morphologies.  Encouragingly, we find that
  galaxy disks are common in moderate-feedback runs, making up
  typically $\sim 50\%$ of all galaxies at the centers of haloes with
  virial mass exceeding $\sim 10^{11} \, M_{\odot}$. The size, stellar
  masses, and circular speeds of simulated galaxies formed in such
  runs have properties in between those of large star-forming disks
  and of compact early-type galaxies at $z=2$. Once the detailed
  abundance and structural properties of these rare objects are well
  established it may be possible to use them to gauge the overall
  efficacy of feedback in the formation of high redshift galaxies.
\end{abstract}


%%%%%%%%% --------------------------------------------- %%%%%%%%%%%%%

\section{Introduction}
\label{sec:intro}

The established paradigm for structure formation offers a clear road
map for galaxy formation. Primordial fluctuations in the dominant cold
dark matter (CDM) component of the Universe grow via gravitational
instability, sweeping baryons into an evolving hierarchy of dark
matter haloes that grow through mergers of preexisting units as well as
through the accretion of material from the intergalactic medium
\citep{White1978}. On galaxy mass scales, baryons caught in a halo are
able to radiate away the gravitational energy gained through the
collapse, sink to the center of the halo, and assemble into the dense
aggregations of gas and stars that we call galaxies \citep{Blumenthal1985}.

The structure and morphology of a galaxy results from the complex
interplay between the time of collapse, the mode of assembly, the
efficiency of cooling, and the rate of transformation of gas into
stars \citep[see, e.g.,][]{SteinmetzNavarro2002}. Where cooling
dominates and outpaces star formation, baryons collect into thin,
rotationally-supported disks \citep{Fall1980}. Stars formed in these
disks inherit these morphological features, but are vulnerable to
swift transformation into dispersion-supported spheroids during
subsequent merger events \citep{Toomre1977}. Disks may re-form if
mergers or accretion bring fresh supplies of cooled gas, making
morphology a constantly evolving rather than an abiding feature of a
galaxy \citep{Cole2000,Robertson2006}.

The galaxy formation scenario driven by gravitational collapse and
radiative losses outlined above is compelling, but incomplete.
Indeed, cooling is so effective at early times that, unless impeded
somehow, most baryons would be turned into stars in early-collapsing
protogalaxies, which would then merge away to form by the present time
a majority of spheroid-dominated remnants, in vehement disagreement
with observations \citep{White1978,Cole1991,White1991}. The problem is
compounded by the fact that, during mergers, cooled gas tends to
transfer its angular momentum to the surrounding dark matter halo. As
a result, even in cases where disks could form, their structural
properties would be at odds with those of spiral galaxies
\citep{Navarro1991,Navarro1995,NavarroSteinmetz1997}.

A gas heating mechanism that prevents runaway cooling and that
regulates the formation of stars in step with mergers and accretion is
widely believed to be the most likely solution to these problems. The
energetic output from evolving stars and supernovae is a natural
candidate. It scales directly with star formation and, in a typical
galaxy, the total energy released by supernovae can be comparable to
the binding energy of the baryons. Thus, if channeled properly,
feedback energy from supernovae may temper the gravitational
deposition of cooled gas into a galaxy and effectively self-regulate
its star formation history \citep{White1991}.

The even standing of gravity, feedback and cooling may thus help
reconcile the observed galaxy population with hierarchical clustering
models, but it comes at the price of complexity: the main structural
properties of a galaxy, such as stellar mass, rotation speed, and
morphology, are then expected to depend on details of its assembly
history, such as the exact timing, geometry and mass spectrum of
accretion events \citep[see,
e.g.,][]{Abadi2003a,Abadi2003b,Meza2003,Governato2007,Zavala2008,
Scannapieco2009,Governato2010}.

Such sensitivity to feedback has held back progress in direct
simulation of the process of galaxy formation. As recent work
demonstrates, different but plausible implementations of feedback
within the {\it same} dark halo lead to galaxies of very different
mass, morphology, dynamics, and star formation history
\citep[see, e.g.,][]{Okamoto2005}. Numerical parameters may thus be tuned to
reproduce some properties of individual galaxies, but at the expense
of wider predictability in the modeling.

These results suggest that further progress in the subject
requires the testing of different feedback schemes on a statistically
significant sample of dark haloes formed with representative assembly
histories. The viability of each feedback implementation may then be
assessed by contrasting the statistics of such samples with
observational constraints such as the stellar mass function,
clustering, color distribution, and scaling laws.

We take a step in this direction here by analyzing a subset of
cosmological N-body/gasdynamical simulations from the OverWhelmingly
Large Simulations ({\small OWLS}) project \citep{Schaye2010}. We present 
results regarding the morphology, stellar mass, and angular
momentum content of galaxies assembled at $z=2$, and compare them with
the few observational constraints available at that epoch. We limit
our analysis to the $z=2$ galaxy population because most
high-resolution {\small OWLS} runs follow volumes too small to be
evolved until $z=0$.  In future papers, we plan to extend this
analysis to the present-day galaxy population using samples drawn from
the closely-related {\small GIMIC} project, designed to follow a few
representative volumes selected from the Millennium Simulation
\citep{Crain2009}.

The paper is organized as follows. In Sec.~\ref{sec:numsim} we present a
short overview of the simulations and feedback models. We then present
our main numerical results in Sec.~\ref{sec:numres} and analyze them
in the context of available observational constraints in
Sec.~\ref{sec:obsdiag}. We end with a brief summary in
Sec.~\ref{sec:conc}.

\section{The Numerical Simulations}
\label{sec:numsim}

\subsection{The OWLS runs}

The {\small OWLS} project consists of a suite of $\sim 50$ different
cosmological N-body/SPH simulations that follow the evolution of dark
matter and baryons in boxes of 25 and 100 $h^{-1}$ Mpc (comoving). Each
box is run many times, varying the numerical implementation of
various aspects of the gas cooling, star formation and feedback
modules \citep[see ][ for further details]{Schaye2010}.

We have selected for our analysis nine $25 \, h^{-1}$ Mpc-box {\small
  OWLS} runs, eight of which explore different feedback
implementations with $512^3$ dark matter and $512^3$ baryonic
particles whilst keeping other subgrid parameters constant, such as
the stellar initial mass function (IMF), the star formation threshold
and its efficiency. The ninth repeats one of the runs, at $8\times$
lower mass resolution (and $2\times$ lower spatial resolution), in
order to provide some guidance regarding the sensitivity of our results
to numerical resolution. 

All simulations assume a standard WMAP-3 $\Lambda$CDM cosmogony, start
at $z_i=127$ and, because of their small box size, they have only been
carried out to $z=2$. We adopt this cosmology for all physical
quantities listed here. We make explicit the dependence on the Hubble
constant, $h$, for simulation parameters, but drop the $h$ dependence and
adopt  $h=0.73$ when comparing with observations.

The high-resolution runs have a comoving gravitational softening lengthscale of
1/25 of the initial mean inter-particle spacing at high redshift. These are
switched later to a fixed physical value so that the softening never
exceeds $0.5 \, h^{-1}$ kpc (physical). The mass per baryonic particle is $\sim 1.4
\times 10^6 \, h^{-1} {\rm M}_\odot$ and $4.5$ times higher for the
dark matter component. All runs assume that the
Universe is reionized at $z=9$ (for H) and at $z=3.5$ (He) by a bath
of energetic photons whose properties evolve as proposed by \citet{Haardt2001}.

Table \ref{tab:simpar} summarizes the most important numerical
parameters of the simulations, as well as the cosmological parameters.

%%%%%%TABLE%%%%%%%%%
\begin{center}
\begin{table}
\caption{Simulation parameters}
\begin{tabular}{|c|c|}
\hline
\hline
$\Omega_{\rm M}$ & 0.238\\
$\Omega_{\rm CDM}$ & 0.1962\\
$\Omega_b$ & 0.0418\\
$\Omega_\Lambda$ & 0.762\\
$\sigma_8$ & 0.74\\
$h$ & 0.73 \\
$n$ &  0.951\\
Reionization redshift & 9 (H), 3.5 (He)\\
Mass per DM particle & $m_p=6.3 \times 10^6 h^{-1} \rm M_\odot$ \\
Mass per baryonic particle & $m_p = 1.4 \times 10^6 h^{-1} \rm M_\odot$\\
Number of particles & $2 \times 512^3$ \\
Box size & $25 \, h^{-1}$ Mpc\\
\hline
\end{tabular}
\label{tab:simpar}
\end{table}
\end{center}
%%%%%%%%%%%%%%%%%%%%

\subsection{Subgrid gas physics}

Baryons are assumed to trace the dark matter distribution at the
initial redshift. Whilst in gaseous form, they are followed
hydrodynamically and are subject to pressure gradients and
shocks. Radiative cooling and heating is implemented following
Wiersma et al. (2009a), which also accounts for the photo-ionisation of
metals due to the UV background.\nocite{Wiersma2009a}

In collapsed structures, gas can cool and sink to the center of these
haloes, where it may reach high overdensities before turning into
stars. With limited numbers of particles, these regions are poorly
resolved and vulnerable to numerical instabilities, such as artificial
clumping and fragmentation. As discussed by \citet{Springel2003},
these shortcomings can be alleviated by adopting, in high-density
regions, a multi-phase description for the gas where the effective
equation of state differs from the simple ideal gas law.  In practice,
we impose a polytropic equation of state (PEOS; $P \propto
\rho^{\gamma}$, with $\gamma=4/3$) on all gas particles whose density
exceeds a critical value of $n_c=0.1$ cm$^{-3}$, the density above
which the gas is expected to be multiphase and unstable to star
formation \citep{Schaye2004}. This choice ensures that the Jeans mass
in high-density regions is independent of $\rho$, effectively
suppressing artificial clumping and reducing the dependence of 
star formation algorithms on numerical resolution \citep{Schaye2008}.

\subsection{Star formation algorithm}

Star formation is implemented as described in detail by
\citet{Schaye2008}. In brief, stars form out of PEOS gas particles
with pressure-dependent parameters chosen to reproduce a
Kennicutt-Schmidt law with index $1.4$ \citep{Kennicutt1998}. We
assume a Chabrier initial mass function \citep{Chabrier2003} in order
to take into account the enrichment and energy injected into the
surroundings of young star particles by the explosion of SNII and SNIa
supernovae.  The energy per supernova explosion is chosen to be
$10^{51}$ ergs.  These events, together with mass loss from
intermediate mass stars, pollute neighboring gas particles with
metals, as described in Wiersma et al. (2009b). We track 11 species
and include them in the computation of the cooling function following
(Wiersma et al 2009a) in all our runs, with the exception of the
``NoF'' model described below. For the latter case, chemical enrichment is
modelled in the same way but is not considered in the computation of
cooling, which instead assumes primordial abundances.
\nocite{Wiersma2009b}

\subsection{Feedback Models}

The runs we analyze here explore alternative feedback implementations
where the total amount of energy injected by supernovae into the
surrounding ISM is kept constant, but the numerical algorithm used to
inject this energy is varied. All runs that include feedback from core
collapse supernova feedback assume a total energy input of $10^{51}$
ergs per solar mass of stars formed, 40\% of which is invested into
driving outflowing winds. The remainder is assumed to be lost to
radiative processes.

\subsubsection{Thermal Feedback}

The simplest possibility, which we label "thermal feedback" (ThF), is
to use the supernova energy to raise the internal energy of the
surrounding gas particles. As reported in earlier work
\citep{Katz1992}, these regions typically have such short cooling
times that the injected energy is quickly radiated away, with little
hydrodynamical effect on the surrounding gas. As a result, thermal
feedback is rather inefficient, and has little effect in regulating
gas cooling and star formation, even though the implementation here follows the
stochastic heating method described in \citet{Schaye2010} and presented in more detail
in Dalla Vecchia et al. (in preparation), which is
more resilient to numerical resolution limitations than the
implementations adopted in earlier work \citep[see also][]{Kay2003}.

\subsubsection{Kinetic Feedback}
A second possibility is to invest part of the feedback energy directly
into gas bulk motions, with the aim of allowing gas to outflow from
regions of active star formation, thus increasing the overall
efficiency of feedback. These wind models are characterized by a
couple of parameters: a ``mass loading'' factor, $\eta$, specifying the
number of gas particles amongst which the injected energy is shared,
and a ``wind velocity'', $v_{\rm w}$, characterizing the kinetic energy of the
outflow. For given energy, $\eta$ and $v_{\rm w}$ are related by a constant
$\eta \, v_{\rm w}^2$. 

For the reference model in OWLS (WF2 in our notation) the wind
velocity $v_{\rm w}=600$ km/s is chosen partly motivated by observation of
local starburst galaxies \citep{Veilleux2005}.  The mass
loading $\eta$ is thus fixed to 2 particles as the integer number that
best reproduces the peak in the cosmic star formation history. This
combination of $v_{\rm w}$ and $\eta$ imply that 40\% of the total energy
liberated by supernovae impacts the kinematics of the surrounding
gaseous medium.  Because all these parameters are highly uncertain, in the
OWLS runs we contrast the results obtained with three different values
of $\eta$=1,2,4 (we refer to these runs as WF1, WF2, and WF4, respectively). 
The wind velocities in each model are adjusted so
that the same amount of energy (40\%) is input in every case (see Table~\ref{tab:fpar}).
Further 
details can be found in Dalla Vecchia \& Schaye (2008) and Schaye et
al. (2010). WF2LR is equivalent to WF2
but run at $8 \times$ poorer mass resolution and $2\times$ poorer
spatial resolution.

As discussed by \citet{Springel2003}, a possible modification that can
enhance feedback efficiency is to temporarily "decouple" the wind
particle(s) hydrodynamically from the surrounding ISM. This
facilitates large-scale galactic outflows and regulates star formation
more effectively by enhancing the removal of gas from active
star-forming regions \citep[see e.g.,][]{DallaVecchia2008}.  The
  criterion for re-coupling particles to the gas is as in Springel \&
  Hernquist, and occurs as soon as either: $a)$ the density has fallen
  to 0.1 $n_c$, where $n_c$ is the density threshold for
  star formation, or $b)$ 50 Myr have elapsed since decoupling. We
label this run WF2Dec.

A further run probes the possibility that the efficiency of feedback
should correlate with the local density of the gas. We explore a model
in which the wind velocity and mass loading are related to the gas
density by $v_{\rm w} \propto \rho^{1/6}$ and $\eta \propto
\rho^{-1/3}$. This guarantees that the wind velocity scales with the 
local gas sound speed ($v_{\rm w} \propto c_s$) given the
aforementioned effective PEOS that holds in
star-forming regions. The $v_{\rm w}$ and $\eta$ relations are
normalized so that, at the gas density corresponding to the star
formation threshold ($n_c = 0.1$ cm$^{-3}$), they match $v_{\rm
  w}=600$ km/s and $\eta=2$ particles, consistent with the WF2 run. We
will refer to this run as WDENS.

\subsubsection{AGN Feedback}

Our next model enhances feedback by adding to the WF2 feedback the
extra energetic input from AGN.  This model, which we refer to as AGN,
for short, follows the numerical procedure introduced by
\citet{Booth2009} and summarized in \citet{Schaye2010}. Seed black
holes with mass $m_{\rm seed}=9 \times 10^{4} M_\odot$ are placed at
the center of all haloes that exceed a threshold virial mass of $4
\times 10^{10} h^{-1}M_\odot$. BHs can then grow by mass accretion and
mergers with other BHs. Gas accretion onto the BH is modelled
according to a modified version of the Bondi-Hoyle-Lyttleton
\citep{BondiHoyle1944,Hoyle1939} formula: $\dot{m_{\rm accr}} =\alpha
\, 4\pi \, G^2 \, m_{\rm BH}^2 \, \rho/(c_s^2+ v^2)^{3/2}$, where
$m_{\rm BH}$ is the mass of the BH, $v$ is the velocity of the BH with
respect to the ambient medium, $c_s$ is the local speed of sound and
$\rho$ the local gas density. $\alpha$ is an extra ``efficiency''
parameter that did not appear in the original versions of the
Bondi-Hoyle formula but was introduced by \citet{Springel2005c}, who
set it to alpha = 100, to account for the finite numerical resolution
and for the fact that the cold, interstellar phase is not explicitly
modeled. In our model this factor is set to unity in the regime that
the physics is modeled correctly, but increases with the ISM density
(i.e., for particles on the PEOS, see Booth \& Schaye (2009) for
further details and discussion).

A fraction $\epsilon_f$ of the total radiated energy due to the mass
accretion onto the BHs is assumed to couple to the surrounding
ISM. This efficiency is set to $\epsilon_f=0.15$ to match local
constraints on the number density \citep[see, e.g.,][]
{Shankar2004,Marconi2004} as well as relations between BHs and
host galaxy properties \citep{Haring2004,Tremaine2002}, 
both at redshift zero. AGN feedback is implemented as a {\it
  thermal} injection of energy (as opposed to the kinetic prescription
used to model the stellar feedback), in the way described in
\citet{Booth2009}. Because it combines the supernova and AGN energetic
outputs, the AGN run is the most effective feedback model tried in our
series.

\subsubsection{No Feedback}

Finally, mainly for comparison purposes we also analyze a run that
follows star formation like in the other implementations but neglects
all energy injection into the ISM due to either supernovae or AGN.  Gas
cooling in this ``no feedback'' model, NoF, adopts the cooling
function of a gas with primordial abundances, but in the absence of
feedback this is only a minor difference that has little impact on the
results. The NoF model stands at the opposite extreme as AGN, allowing
for unimpeded transformation of gas into stars in regions able to
collapse and condense into galaxies. Although unrealistic as a galaxy
formation model, it serves to provide a useful framework where the
relative importance of feedback effects may be gauged and understood.

Table~\ref{tab:fpar} summarizes the relevant parameters of each
feedback implementation. For ease of reference, we also quote in each
case, the name used to label each simulation by \citet{Schaye2010}.

%%%%%%TABLE%%%%%%%%%
\begin{center}
\begin{table} 
  \caption{Parameters of the different feedback models
    probed in each run. First and second columns list the short name
    (used throughout this paper) and the name originally used in
    \citet{Schaye2010}, respectively. The third and fourth columns list
    the mass loading ($\eta$) and wind velocity
    ($v_{\rm w}$) parameters of each model. The WF2Dec is the only
    model where wind gas particles are temporarily kinematically
    decoupled from the surrounding gas. This aids the removal of gas
    from galaxies and results in increased feedback efficiency.
    The characteristic density $n_{\rm w}$ used for scaling the
    WDENS wind parameters is that corresponding to
    the star formation threshold: $n_c = 0.1$ cm$^{-3}$.}
\begin{tabular}{|c|c|c|c|}
\hline
Short name & OWLS name & $\eta$ & $v_{\rm w}$\\
& & [particles] & [km/s]\\
\hline
\hline
NoF & NOSN$\_$NOZCOOL & -- & --\\
 ThF & WTHERMAL  & -- & -- \\
WF4 & WML4V424 & 4 & 424 \\
WF2 & REF & 2 & 600  \\
WF1 & WML1V848 & 1 & 848 \\
WF2Dec & WHYDRODEC & 2 & 600 \\
WDENS & WDENS & $ 2 (n/n_{\rm w})^{-1/3} $ & $ 600(n/n_{\rm w})^{1/6}$ \\
AGN & AGN & 2 & 600\\
\hline
\end{tabular}
\label{tab:fpar}
\end{table}
\end{center}
%%%%%%%%%%%%%%%%%%%%

%%%%%%%%%%%%%%%%%%%%%%%%%%
\begin{center}
\begin{figure*}
\includegraphics[width=0.75\linewidth,clip]{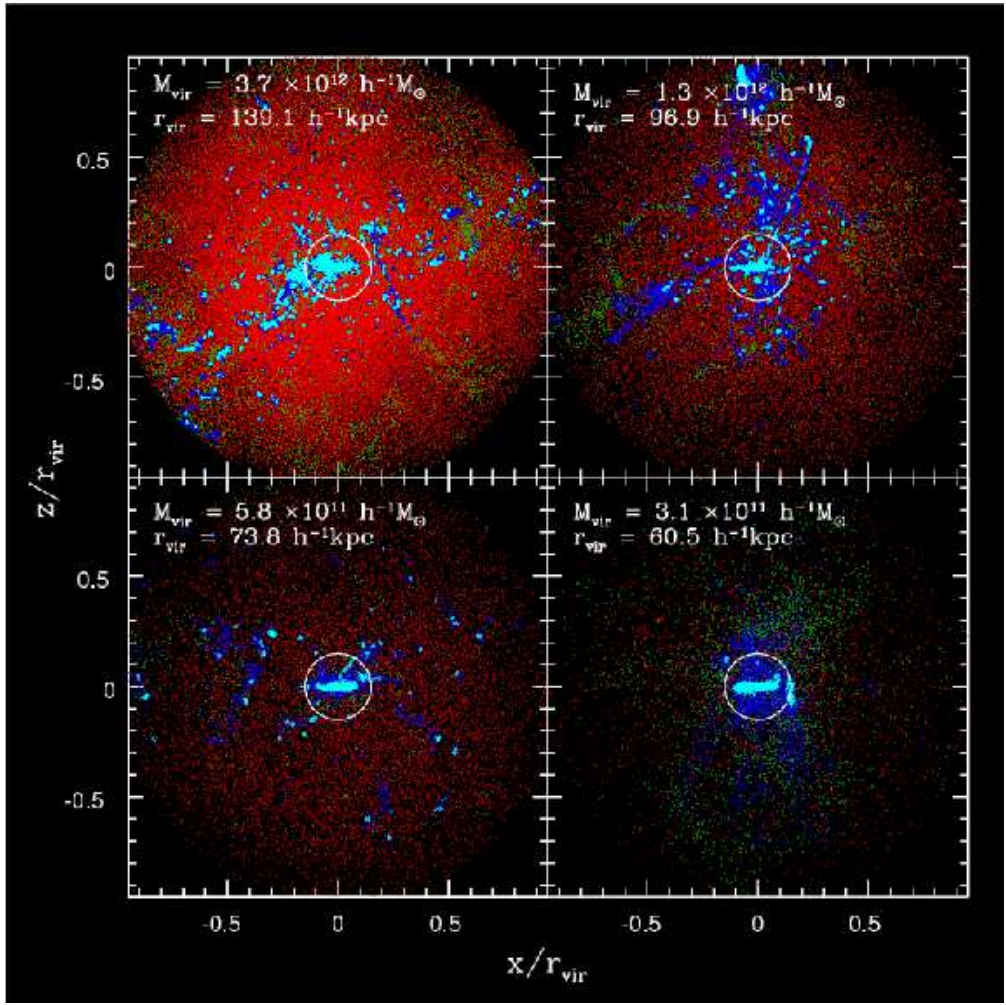}
\caption{
Gas particles within the virial radius of four WF2 haloes spanning the
mass range of systems selected for analysis. The virial mass and
radius are given in the label of each panel. Gas particles are colored
according to temperature: red, green, and blue correspond to particles
in the hot, warm, and cold phases, respectively. Hot particles are
those with $T> (1/4) T_{\rm vir}$, where $T_{\rm vir}=35.9 (V_{\rm
vir}/$ km s$^{-1})^2$ K is the virial temperature of the halo. Cold
particles are those with $T< 3 \times 10^4$ K. Warm are those with
intermediate temperatures. Cyan particles denote dense, star-forming
gas in the PEOS phase. Particles are plotted sequentially in order of descending
temperature, so colder particles may occult hotter ones in regions of
high density. Small circles show the radius, $r_{\rm gal}=0.15 \, r_{\rm
  vir}$, used to define the central galaxy.
}
\label{fig:GasRvir}
\end{figure*}
\end{center}
%%%%%%%%%%%%%%%%%%%%%%%%%%

%%%%%%%%%%%%%%%%%%%%%%%%%%
\begin{center}
\begin{figure*}
\includegraphics[width=0.75\linewidth,clip]{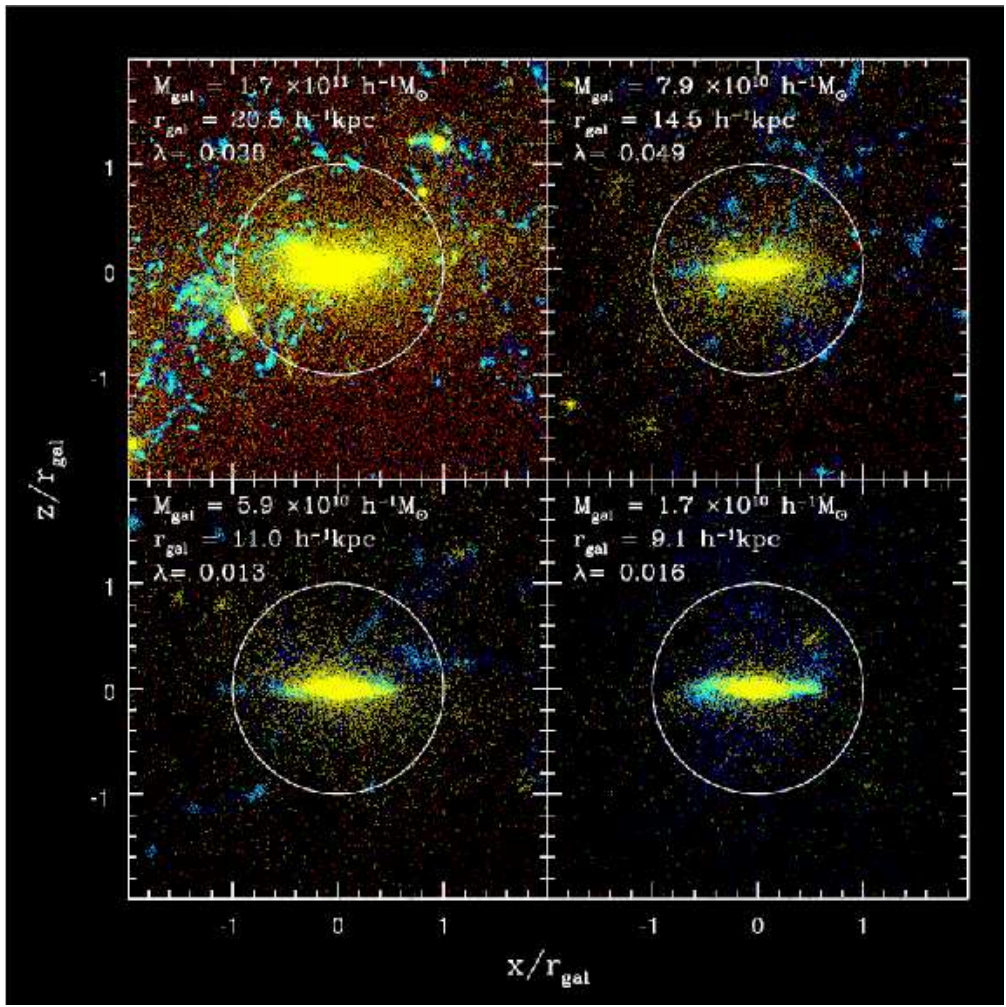}
\caption{
Zoomed-in view of the galaxies at the centers of the haloes shown in Fig.~\ref{fig:GasRvir}. Colors
are as described in the caption of that figure, except that yellow now
denotes ``star'' particles. The circles show the galaxy ``radius'', $r_{\rm gal}=0.15 \,
r_{\rm vir}$. Each box has been rotated so that the spin axis of the
PEOS gas is aligned with the $z$ axis of each panel. This
``edge-on'' projection emphasizes the presence of disk-like structures
in all four haloes. Besides $r_{\rm gal}$, labels in each panel specify
the baryonic mass of the galaxy, $M_{\rm gal}$, and the spin parameter
of the surrounding halo, $\lambda$. 
}
\label{fig:GasStarsRgal}
\end{figure*}
\end{center}
%%%%%%%%%%%%%%%%%%%%%%%%%%

%%%%%%%%%%%%%%%%%%%%%%%%%%
\begin{center}
\begin{figure*}
\includegraphics[width=0.475\linewidth,clip]{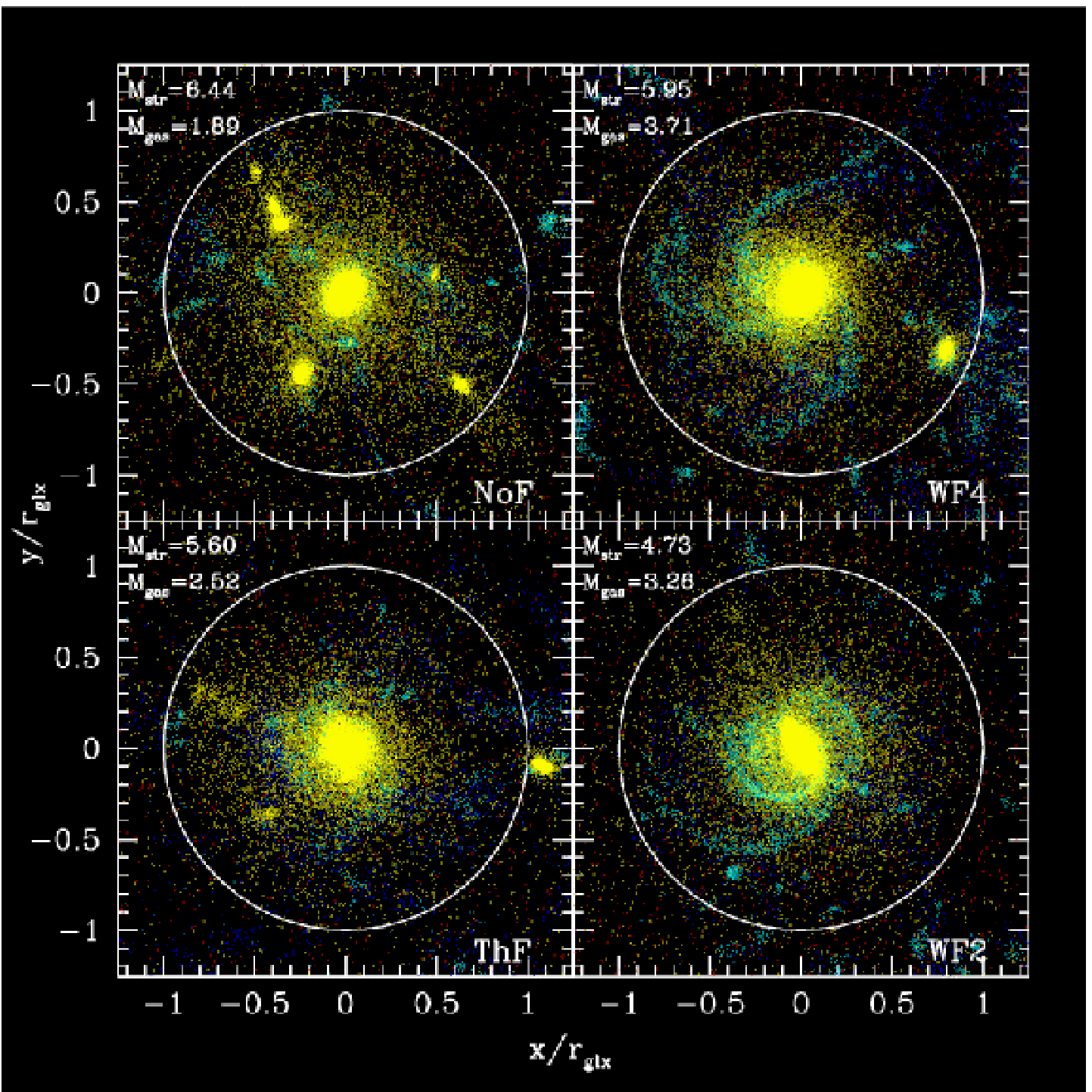}
\hspace{0.4cm}
\includegraphics[width=0.475\linewidth,clip]{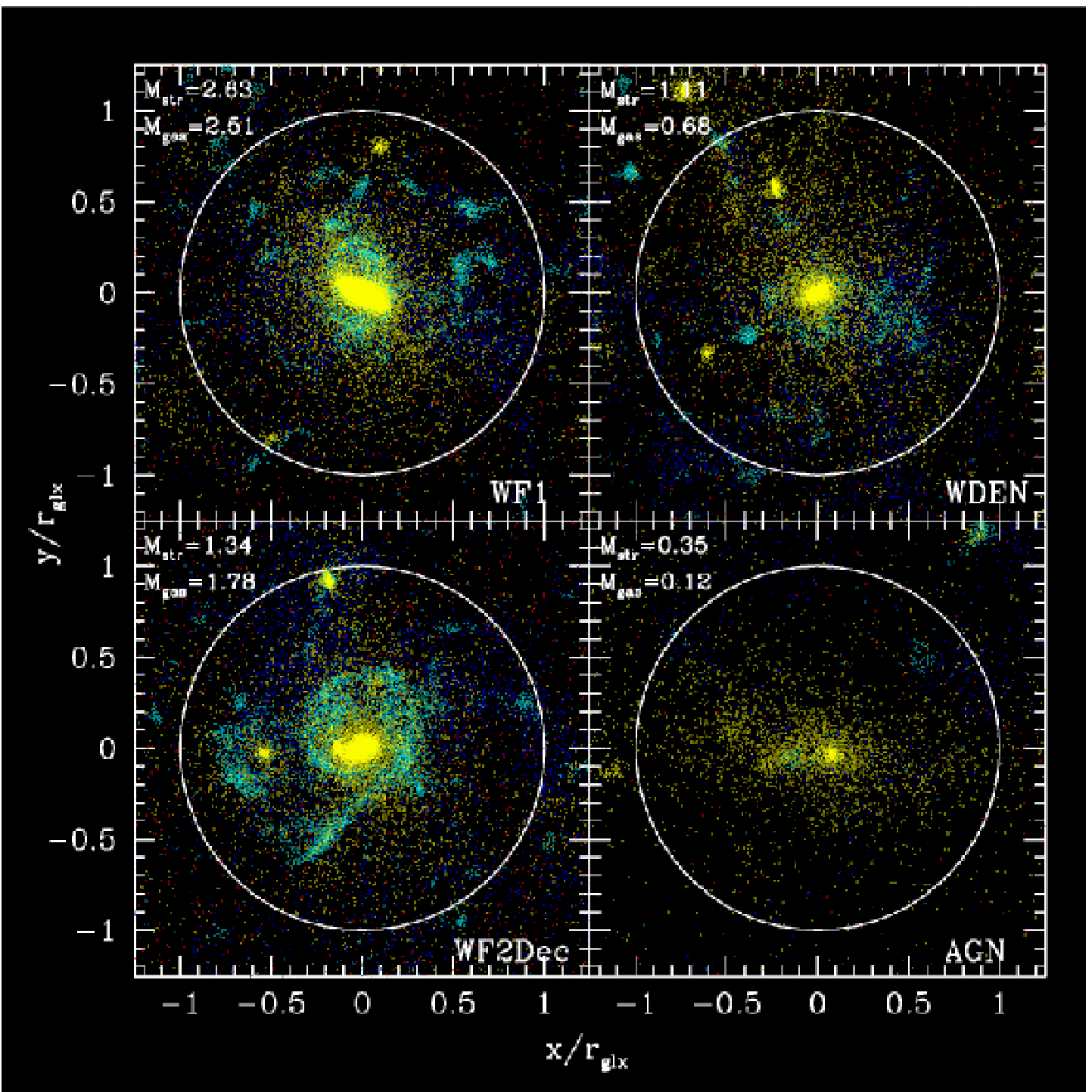}
\caption{ Face-on view of the central galaxy formed at the center of
  an $M_{\rm vir}=1.2 \times 10^{12} \, h^{-1} \, M_{\odot}$ halo. All
  panels correspond to the {\it same} halo, but in runs with different
  feedback implementations, as labelled in the bottom right of each panel. Only baryonic
  particles are shown. Colors indicate gas temperature, classified as
  hot (red), warm (green), cold (blue) and star-forming (cyan). See
  the caption to Fig.~\ref{fig:GasRvir} for details. Yellow dots
  correspond to ``star'' particles. The circle in each panel indicate
  the radius used to define the central galaxy, $r_{\rm gal}$. Each
  galaxy has been rotated so that it is seen ``face on''; i.e., the
  angular momentum of the PEOS gas is aligned with the line of sight
  of the projection. The mass in stars and gas within $r_{\rm gal}$ is
  labelled in each panel (units are $10^{10} \, h^{-1} M_\odot$).  }
\label{fig:FaceOnGx}
\end{figure*}
\end{center}
%%%%%%%%%%%%%%%%%%%%%%%%%%

%%%%%%%%%%%%%%%%%%%%%%%%%%
\begin{center}
\begin{figure*}
\includegraphics[width=0.475\linewidth,clip]{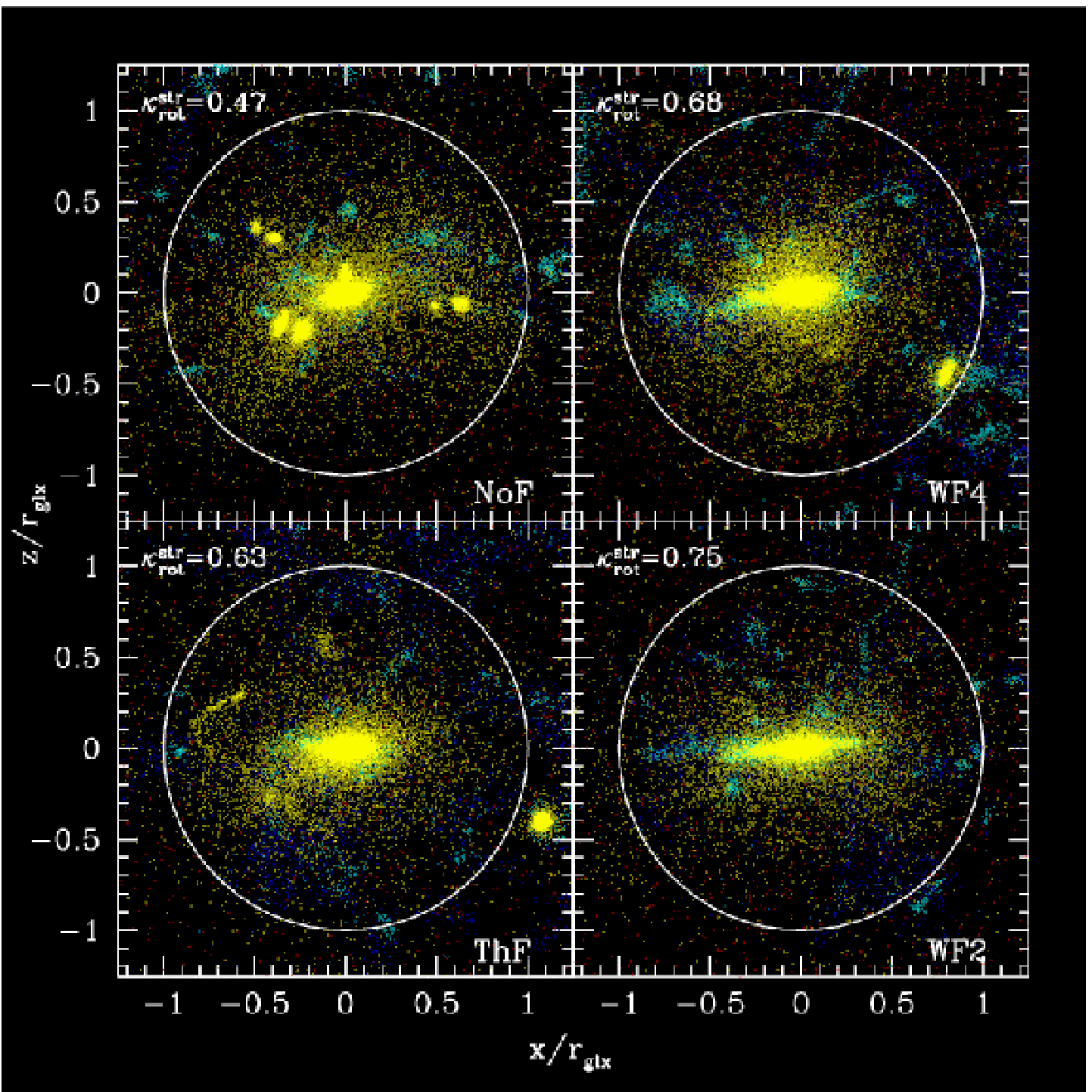}
\hspace{0.4cm}
\includegraphics[width=0.475\linewidth,clip]{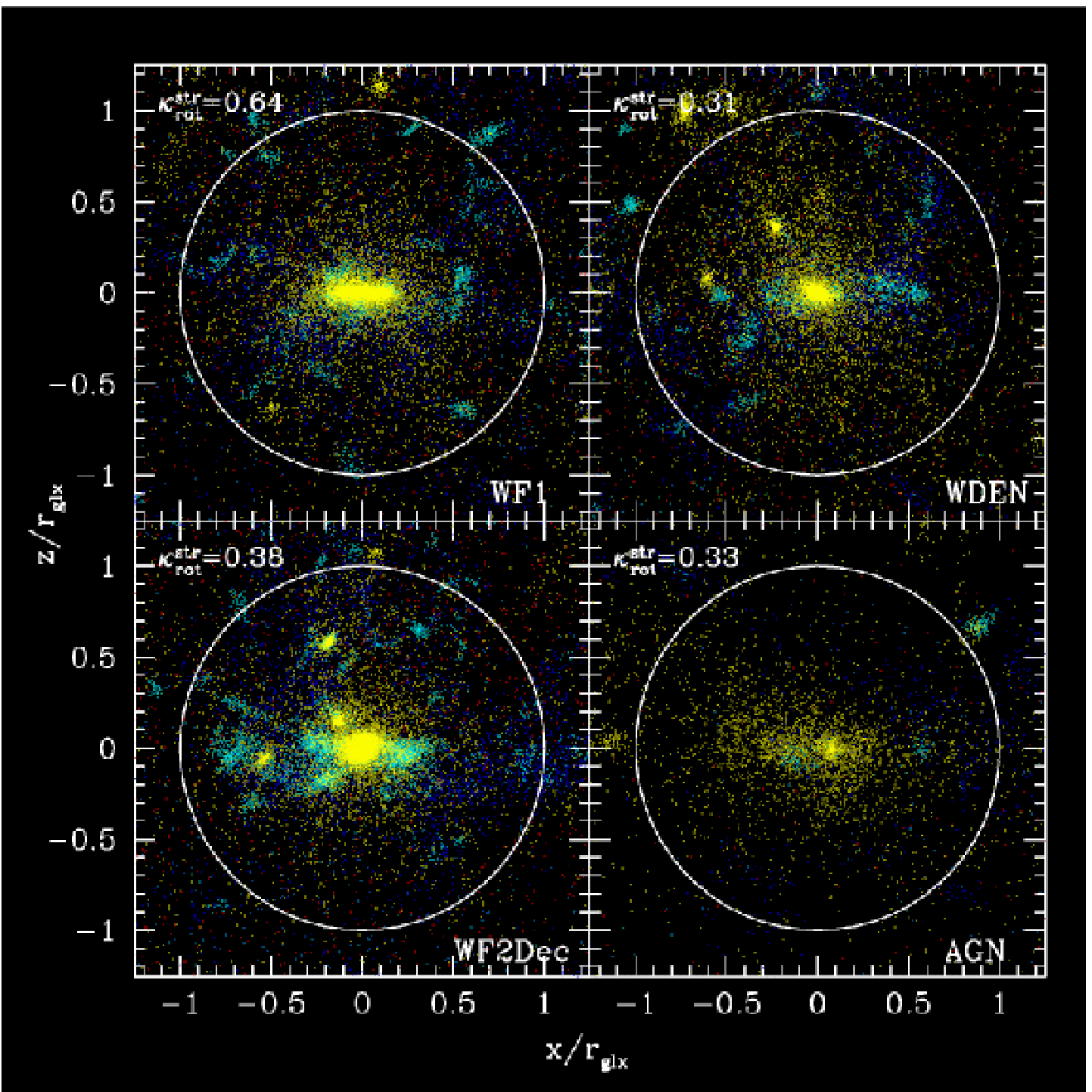}
\caption{ Same as Fig.~\ref{fig:FaceOnGx}, but each galaxy has been
  rotated so that it is seen ``edge-on''. Labels in each panel give the
  fraction of kinetic energy of the stellar component in ordered
  rotation.  }
\label{fig:EdgeOnGx}
\end{figure*}
\end{center}
%%%%%%%%%%%%%%%%%%%%%%%%%%

\section{Numerical Results}
\label{sec:numres}

\subsection{The halo sample}

Our sample consists of all galaxies at the centers of haloes with
virial mass\footnote{Virial values are measured at or within the virial radius,
$r_{\rm vir}$, of a halo, defined as the radius where the mean inner
density exceeds the critical density of the universe by a factor
$\Delta_{\rm vir}(z)=18\pi^2+82f(z)-39f(z)^2$. Here $f(z)=[\Omega_{\rm
M}(1+z)^3/(\Omega_{\rm M}(1+z)^3+\Omega_\Lambda))]-1$ and $\Omega_{\rm
M}=\Omega_{\rm CDM}+\Omega_{\rm
bar}$ \citep{bryanandnorman98}. $\Delta_{\rm vir}\sim 170$ at $z=2$ for
our choice of cosmological parameters.}
$M_{\rm vir} > 10^{11} \, h^{-1}\, M_\odot$. There are about
$150$ haloes at $z=2$ in each $25 \, h^{-1}$ Mpc-box
{\small OWLS} run with masses between $10^{11} \,
h^{-1} \, M_\odot< M_{\rm vir} < 3\times 10^{12} h^{-1} \,
M_\odot$. The median of the sample is $M_{\rm vir} \sim 1.8\times 
10^{11}  h^{-1} \, M_\odot$. Halos are identified by the 
substructure finding algorithm {\small SUBFIND} \citep{Springel2001a, 
Dolag2009} and centers are defined by the minimum of the potential. 
All runs use the same initial conditions, and therefore the number 
(and identity) of haloes selected for analysis is roughly the same 
in each simulation.

We begin with an overview of the properties of the gaseous component
within the virial radius and its halo mass dependence
(Sec.~\ref{SecGasRvir}), and follow on with a description of the
properties of the stellar component of the central galaxy
(Sec.~\ref{SecCentrGx}). We discuss the link between feedback and
morphology in Sec.~\ref{SecFbMorph}, and compare the number of massive
galaxies in various runs in Sec.~\ref{SecMassGx}.  We end this section
by discussing the mass and angular momentum of central galaxies, as
well as their dependence on feedback (Secs.~\ref{SecMGx} and
~\ref{SecJGx}), before proceeding to compare these results with
observations.

\subsection{Gas within the virial radius}
\label{SecGasRvir}

Fig.~\ref{fig:GasRvir} illustrates the distribution of gas within the
virial radius in four haloes selected from the WF2 run at
$z=2$. Each panel corresponds to haloes differing by consecutive
factors of two in virial mass. The box size in each panel has been
adjusted to the virial radius of each halo. Only gas particles within
the virial radius are shown, and have been colored according to their
density/temperature.

Red particles are those with temperatures exceeding $(1/4) \, T_{\rm
  vir}$, where $T_{\rm vir}=35.9 \, (V_{\rm vir}/$km s$^{-1})^2$ K is
the virial temperature of a halo ($V_{\rm vir}$ is the circular
velocity at $r_{\rm vir}$). Gas particles in this ``hot phase'' are
all found in a low-density, largely pressure-supported atmosphere that
fills the halo out to the virial radius. The virial temperature is
$\sim 10^6$ K for haloes with $V_{\rm vir} \sim 170$ km/s, about the
median virial velocity range spanned by our sample.  

The fraction of gas in the hot phase increases with halo mass; it
makes up $68\%$ of all the gas within $r_{\rm vir}$ in the most
massive halo but only $21\%$ in the least massive system shown in
Fig.~\ref{fig:GasRvir} . This is a result of the steady increase in
cooling time with increasing halo mass, which favors the formation of
a hot tenuous gas atmosphere in massive systems.

Particles in green are those in the ``warm'' phase, which we define as
those satisfying $3\times 10^4$K$<T< (1/4) \, T_{\rm vir}$. These are
particles at moderate overdensities, and make up a small fraction of
all the gas within $r_{\rm vir}$; from $\sim 7\%$ in the most massive halo
to $\sim 15\%$ in the least massive one. This gas typically traces
material accreted relatively recently, which has yet to be pressurized
by shocks, or material ejected during accretion events in "tidal
tails" that expand and cool as they recede from the center. Because
accretion occurs frequently through filaments, and tidal tails are
likewise highly asymmetric, the warm component distribution is
non-uniform, with discernible large-scale features suggestive of
recent mergers and accretion events.

Cold ($T<3\times 10^4$ K) gas of moderate density ($n < n_c =0.1$
cm$^{-3}$) is shown in blue, and is rather clumpy in
appearance. Large-scale features similar to those noted for the warm
component are also visible here, suggesting that this is also mostly
gas recently accreted or affected by accretion events. In terms of
mass, this component is negligible ($\sim 5\%$) in the $3.7 \times
10^{12} \, h^{-1} \, M_{\odot}$ halo but increases in importance with
decreasing halo mass. Indeed, it makes up $\sim 30\%$ of all the gas
in the $3.1\times 10^{11} \, M_\odot$ halo shown in
Fig.~\ref{fig:GasRvir}.

The star-forming gaseous component is, by definition, the densest
($n > n_c = 0.1$ cm$^{-3}$), and is shown in cyan in
Fig.~\ref{fig:GasRvir}. Most of this gas is at the bottom of the
potential well of the main halo and of its substructure haloes, and
makes up between $20$ and $30\%$ of the gas within $r_{\rm vir}$, with
little dependence on halo mass.

The generally strong halo mass dependence of the various gaseous phases
highlights the different modes of accretion that shape the evolution of
a central galaxy. In massive haloes galaxies grow by accreting cooled
material from the surrounding reservoir of hot gas, whereas in low
mass haloes the gas is likely to flow virtually unimpeded to the
central regions, where it may be swiftly accreted into the central
galaxy \citep{White1991,Keres2005,Dekel2006,Birnboim2007,Keres2009,Brooks2009}. 
These different accretion modes highlight the complex assembly
history of a galaxy, a complexity that is further compounded by the
effects of feedback that we discuss below.

%%%%%%%%%%%%%%%%%%%%%%%%%%
\begin{center}
\begin{figure}
\includegraphics[width=84mm]{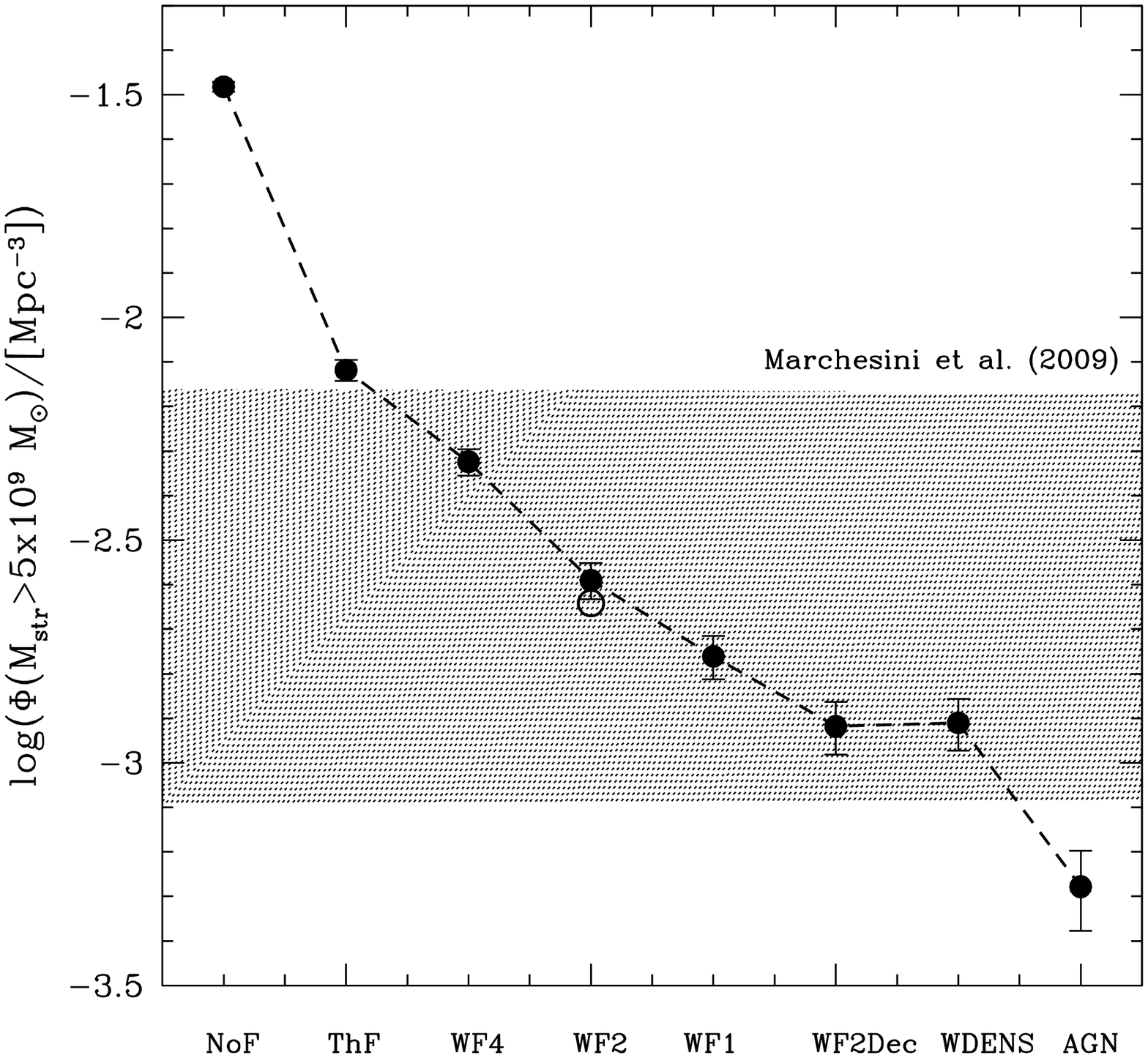}
\caption{ Number density of galaxies with stellar mass exceeding $5
  \times 10^9 \, M_{\odot}$, shown for the various feedback
  implementations explored in this paper. Runs are labeled as in
  Table~\ref{tab:fpar}, and are listed in the abscissa roughly in
  order of increasing feedback efficiency. The open circle correspond
  to the WF2 low resolution run WF2LR. ``Error bars'' denote $\sqrt N$
  uncertainties corresponding to the number of systems in our
  computational box. The shaded area outline observational constraints
  from estimates of the galaxy stellar mass function at $z=2$, as
  compiled by \citet{Marchesini2009}. Note the strong decline in the
  number of massive galaxies as a function of increasing feedback
  efficiency. }
\label{fig:NumDens}
\end{figure}
\end{center}
%%%%%%%%%%%%%%%%%%%%%%%%%%

\subsection{Central galaxies}
\label{SecCentrGx}

Fig.~\ref{fig:GasStarsRgal} shows a zoomed-in view of the four WF2
haloes depicted in Fig.~\ref{fig:GasRvir}, including the stellar
component, which is shown in yellow. The circle centered on the main
galaxy indicates the radius, $r_{\rm gal}=0.15 \, r_{\rm vir}$, that
we use to define the central galaxy inhabiting each halo. As is clear
from the figure, this definition includes virtually all stars and
dense gas obviously associated with the galaxy.

It also emphasizes the halo mass dependence of the various phases in
which baryons may flow into the central galaxy. As discussed above,
whereas galaxies in low mass haloes grow through the smooth accretion
of cold gas, a fair fraction of the star forming gas in the most
massive systems include ``clouds'' that condense out of the hot and
warm phases. Little star formation happens in these clouds, however,
since their typical densities are well below those reached in the main
body of the galaxy.

Gas turns swiftly into stars once it settles into a dense, thin,
rotationally supported disk in the central galaxy. In systems that
avoid major mergers, the stellar component inherits the disk-like
structure of the gaseous component. All 4 galaxies shown in
Fig.~\ref{fig:GasStarsRgal} sport well-defined stellar disks, which
have been rotated to be seen ``edge-on'' in this figure. Disks of gas
and stars are indeed quite common in the WF2 run that we have chosen
to illustrate the main general features of our simulated galaxies.

\subsection{Feedback and morphology}
\label{SecFbMorph}

Varying the feedback implementation has a dramatic effect on the
properties of central galaxies. We illustrate this in
Figs.~\ref{fig:FaceOnGx} and ~\ref{fig:EdgeOnGx}, where we show, for
the {\it same} dark matter halo, how the appearance of its central
galaxy varies with feedback. Although the assembly history of the dark
halo is identical in all cases, differences in feedback lead to
drastic variations in the stellar mass, gaseous content, and
morphology of the central galaxy.

When feedback is inefficient, such as in the ThF and WF4 runs, a
stellar disk is clearly present, but its mass is small compared with
that of the spheroidal component.  This is because most stars form in
early collapsing protogalaxies which are later stirred into a
spheroidal component when these subsystems coalesce to form the final
galaxy. The extreme case is NoF, where the absence of feedback allows
for early and highly efficient star formation that converts most of
the available gas into stars. The large number of satellites seen
around the NoF central galaxy is also a result of the lack of
feedback. This preserves star formation even in small subhaloes, where
modestly energetic feedback might lead to drastic changes in the
availability of star formation fuel and in the total mass of stars
formed.

When feedback effects are strong, such as in the WF2Dec, WDENS, and
AGN runs, fewer stars form since the gas is constantly pushed out of
star-forming galaxies by outflows. These outflows also disrupt the
smooth settling of gas into disks and its gradual transformation into
stars. In the most extreme case (AGN), the gas outflows are so violent
that there is little gas left in the central galaxy. In none of these
cases do central galaxies have an extended and easily recognizable
stellar disk component.

As may be seen from Fig.~\ref{fig:EdgeOnGx}, more moderate feedback
implementations, such as WF2 and WF1, yield systems with a
well-defined stellar disk, and a gas/stellar mass fraction of roughly
$1$:$1$.

This impression is corroborated quantitatively by the fraction of {\it
  stellar} kinetic energy in ordered rotation:

\begin{equation}
 {\kappa_{\rm rot}^{\rm Star}=K_{\rm rot}/K; \phantom{ii}\rm with\phantom{ii} K_{\rm rot}=\sum (1/2) m (j_z/R)^2}
\end{equation}
Here, $m$
is the mass of a star particle; $j_z$ is the z-component of the
specific angular momentum, assuming that the z-axis coincides with the
angular momentum vector of the galaxy, and $R$ is the (cylindrical)
distance to the z-axis. $\kappa_{\rm rot}$ is listed in each panel of
Fig.~\ref{fig:EdgeOnGx} for the stellar component: it is highest
for WF2, and minimum for AGN.

\subsection{Feedback and massive galaxies}
\label{SecMassGx}

A robust way of assessing the effectiveness of the various feedback
implementations explored in these runs is to compute the abundance of
massive galaxies that each predicts. Because the total amount of stars
formed decreases as the feedback becomes more effective, the abundance
of massive galaxies is expected to depend sensitively on
feedback. This is shown in Fig.~\ref{fig:NumDens}, where we plot, for
each implementation, the number of galaxies (per unit volume) with
stellar masses exceeding $5 \times 10^9 \, M_{\odot}$.  This mass
threshold is chosen to roughly coincide with $10,000$ baryonic
particles

The runs in Fig.~\ref{fig:NumDens} are ranked, from left to right, in
order of decreasing number of massive galaxies (i.e., increasing
feedback efficiency).  This figure confirms the strong effect of
feedback on the abundance of massive galaxies.  For example, the AGN
run has $\sim 16$ times fewer such galaxies than the run with only
thermal feedback, ThF, and $\sim 60$ fewer than NoF, the model without
feedback energy injection.

Not only does the total amount of feedback energy matter, but also the
manner in which it is injected. Indeed, large differences are also
obtained for models with the {\it same} feedback strength (as measured by
the total feedback energy per unit stellar mass formed), but that
differ in the combinations of mass loading and wind velocities: ThF, WF4,
WF2, WF1, WF2Dec and WDENS all assume that $40\%$ of the available
supernova energy is invested into winds, yet their predictions for
the number of bright galaxies differ by up to a factor of $\sim 4$.

The results do not seem to depend dramatically on numerical
resolution, as shown by the good agreement between the WF2 and WF2LR
runs (the latter is shown with an open symbol in
Fig.~\ref{fig:NumDens}). Reducing the number of particles by a factor
of eight (as in WF2LR) brings down the number of $M_{\rm gal}>5\times 10^{9} \,
M_{\odot}$ galaxies in the box from $133$ to $123$. The trends shown
in Fig.~\ref{fig:NumDens} are therefore unlikely to be an artifact of
limited numerical resolution.

The shaded band in Fig.~\ref{fig:NumDens} indicates the expected
number of massive galaxies, taken from \citet{Marchesini2009}, after
interpolating their fits to $z=2$ and correcting volume elements to
account for the different cosmology assumed in their work. The band
aims to represent uncertainties due to photometric redshift
inaccuracies, cosmic variance and systematics in the modeling. Notice
that the bright end of the luminosity function traces the abundance of
the most massive objects present at $z=2$ and is, as such,
particularly sensitive to the adopted cosmological parameters.

Given these large uncertainties, it would be premature to use
Fig.~\ref{fig:NumDens} to rule in or out any particular
implementation of feedback but, as the data improve, it might be
useful to revisit this issue to learn which feedback modeling
procedure is favored or disfavored by the data. For clarity, many of the
plots in the analysis that follows will focus on 4 cases that
span the full range of feedback strength shown in
Fig.~\ref{fig:NumDens}: i.e., NoF, WF2, WF2Dec, and AGN. 

We end by noting that different observational diagnostics, such as the
specific star formation rate or the gas content as a function of
galaxy luminosity, could be used to provide further constraints on the
viability of each feedback model. We plan to present a detailed
analysis along these lines in a future paper (see Haas et al., {\it in
  preparation}).

%%%%%%%%%%%%%%%%%%%%%%%%%%
\begin{center}
\begin{figure*}
\includegraphics[width=0.475\linewidth,clip]{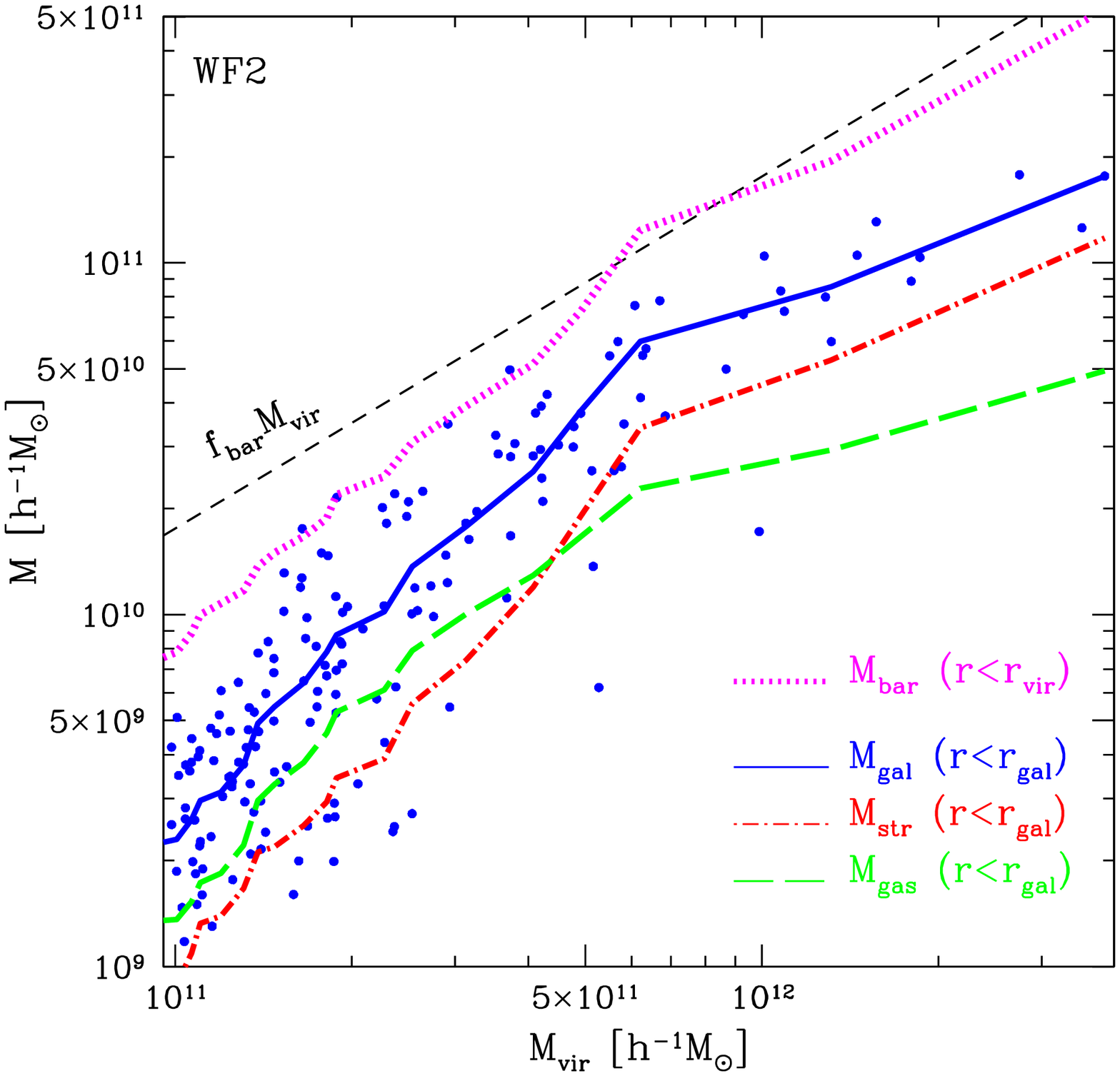}
\hspace{0.4cm}
\includegraphics[width=0.475\linewidth,clip]{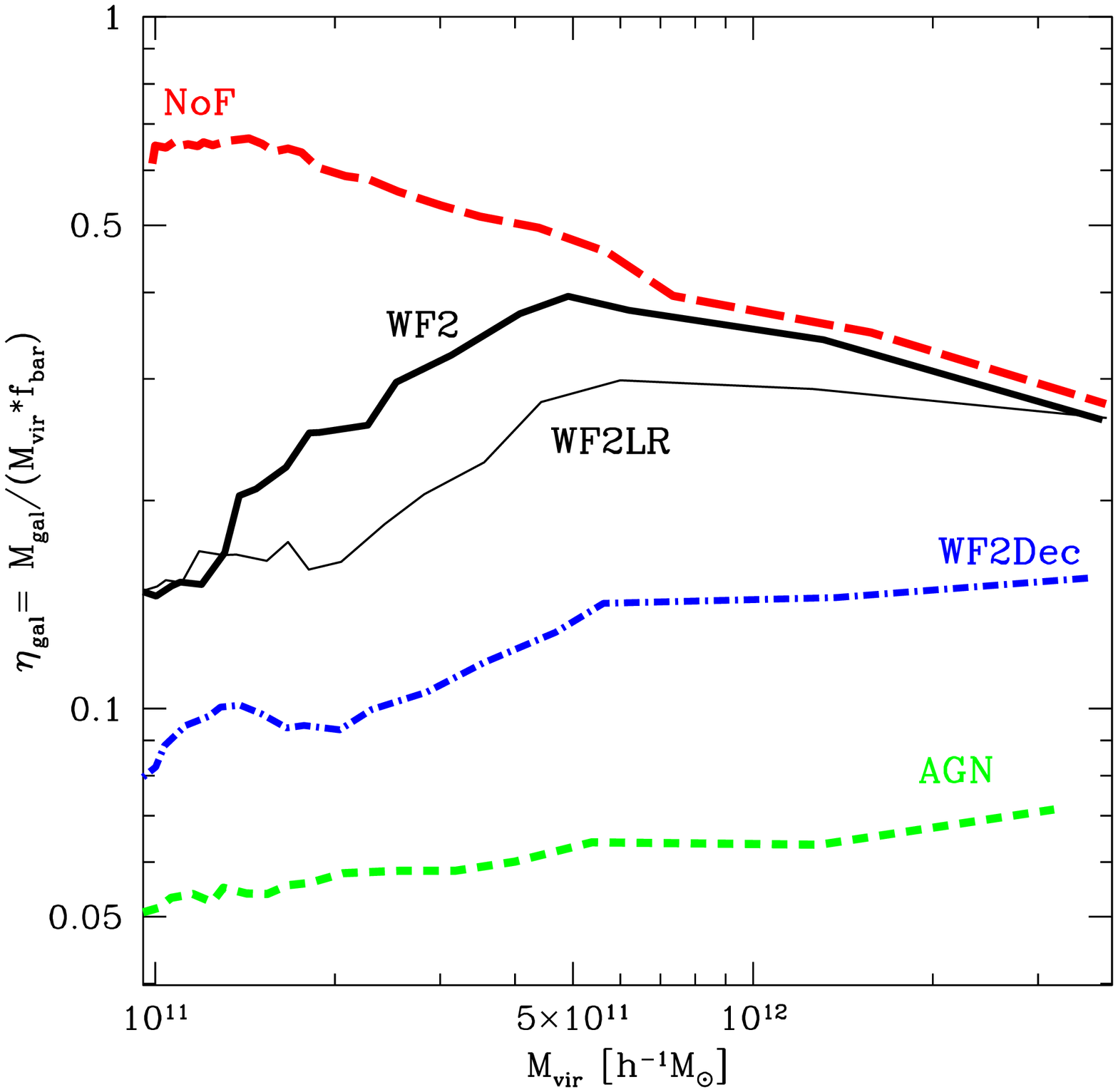}
\caption{{\it Left:} Central galaxy mass as a function of virial mass
  for all haloes selected in the WF2 run. The top dashed line indicates
  the mass in baryons in each halo corresponding to the universal
  baryon fraction, $f_{\rm bar}=\Omega_b/\Omega_M=0.175$, adopted in
  the simulations. Dots correspond to the baryon mass of the central
  galaxy (i.e.; within $r_{\rm gal}$); the thick solid (blue) curve
  tracks its median as a function of $M_{\rm vir}$. The two bottom
  curves track the median for the stellar (red, dot-dashed) and
  gaseous (green, dashed) mass within $r_{\rm gal}$.  The thick dotted
  (magenta) line shows the median of the total mass in baryons within
  the virial radius, $r_{\rm vir}$. {\it Right:} Galaxy formation
  ``efficiency'', $\eta_{\rm gal}=M_{\rm gal}/(f_{\rm bar} M_{\rm
    vir})$, as a function of halo virial mass for the various
  runs. For clarity, only the results corresponding to 4 selected runs
  are shown, spanning the effective range of feedback strength, from
  the ``no feedback'' (NoF) case to the AGN case, where feedback
  effects are maximal. Cases not shown fall between these two
  extremes. The scatter around each curve is large, typically
  $\sim 0.19$ dex rms. The thin line labelled WF2LR has the same physics
  are WF2 but $8 \times$ poorer mass resolution and $2 \times$ lower
  spatial resolution. Note that the galaxy formation efficiency is, on
  average, very sensitive to feedback, but only weakly dependent on
  halo mass, at least for the range of masses considered here.}
\label{fig:BarM}
\end{figure*}
\end{center}
%%%%%%%%%%%%%%%%%%%%%%%%%%
\subsection{Galaxy masses}
\label{SecMGx}

The stellar and gaseous masses of galaxies assembled at the centers of
dark matter haloes are determined largely by the virial masses of the
systems, modulated by the efficiency of radiative cooling and the
regulating effects of feedback. We show this in the left panel of
Fig.~\ref{fig:BarM} for all galaxies selected from the WF2 run. The
dots in the figure correspond to $M_{\rm gal}$, the total baryon mass
within the radius, $r_{\rm gal}$, used to define the central galaxy;
the solid line traces the median as a function of $M_{\rm vir}$. As
expected, the central galaxy mass correlates well with $M_{\rm vir}$,
albeit with fairly large scatter (the global rms about the median trend is
$\sim 0.19$ dex).

The top dashed line in this panel indicates the mass, $f_{\rm
  bar}M_{\rm vir}$, galaxies would have if all baryons in the halo
have assembled at the center (the universal baryon fraction is
$f_{\rm bar}=\Omega_b/\Omega_{\rm M}=0.175$). The thick dotted magenta line
shows the median baryon mass within $r_{\rm vir}$ as a function of
halo mass. This shows that massive systems have retained all baryons within the
virial radius, but also that the effects of feedback are clear at the low mass end:
$10^{11} \, h^{-1} \, M_\odot$ haloes have only retained about
half of their baryons within the virial radius. Of those, only one
third or so have collected in the central galaxy.

Thus, the ``efficiency'' of galaxy formation, as measured by the mass
of the galaxy expressed in units of the total baryon mass
corresponding to its halo, $\eta_{\rm gal}=M_{\rm gal}/(f_{\rm bar}\,
M_{\rm vir})$, increases steadily with halo mass, from $\sim 10\%$ in
$10^{11} \, h^{-1} M_{\odot}$ haloes to a maximum of roughly $40\%$
for $M_{\rm vir} \sim 5 \times 10^{11} \, h^{-1} M_\odot$. There is
also indication that the efficiency decreases in more massive
systems, to roughly $\sim 30\%$ in the most massive haloes.

These trends (i.e., low galaxy formation efficiency in low and high
mass haloes) are qualitatively in line with what is required to
reconcile the shape of the galaxy luminosity function with the dark
matter halo mass function \citep[see,
e.g.,][]{Yang2005,Conroy2009,Guo2010}. Feedback is the main mechanism
responsible for reducing efficiency in low-mass haloes. Together
with long cooling timescales, it also helps prevent the formation of too
massive galaxies in high-mass haloes.

Although the trends seem qualitatively correct, it remains to be seen 
whether a model like WF2, evolved to $z=0$, is able to satisfy the 
stringent constraints placed by the stellar mass function in the local 
Universe. Indeed, the recent estimate of \citet{Li2009} suggest 
that only $3.5\%$ of all baryons in the Universe are today locked up 
in stars, and \citet{McCarthy2009} argue that supernova feedback alone
is not enough to ensure such a low efficiency of transformation of
baryons into stars. Since the runs we analyze here have only been evolved 
to $z=2$, we are unable to address this issue in a conclusive manner, 
but we plan to return to it when extending the present analysis to 
the {\small GIMIC} simulations.

\subsubsection{Feedback dependence}

The right panel of Fig.~\ref{fig:BarM} shows that the overall
efficiency of galaxy formation is quite sensitive to feedback. Each
curve here tracks the median trend of $\eta_{\rm gal}$ with $M_{\rm
  vir}$ for different runs. As expected from the discussion of
Fig.~\ref{fig:NumDens}, $\eta_{\rm gal}$ is highest for NoF and lowest
for AGN, with more moderate results for WF2 and WF2Dec, as well as the
other runs, which are omitted from this panel for clarity.

Clearly, not only the total feedback energy input, but also the
details of its implementation can affect dramatically the galaxy
formation efficiency. Central galaxies in the NoF case can be up to 
$10\times$ more massive than in the AGN run. WF2 galaxies are a factor
of two to three more massive than those formed in WF2Dec, although the
only difference between these two runs is the choice to ``decouple''
hydrodynamically the supernova-driven winds in the latter. The
reasonable agreement (given the large scatter) between WF2 and WF2LR suggests that this result
is not unduly influenced by numerical resolution.

Although the average galaxy formation efficiency depends strongly on
feedback, its dependence on halo mass is weak: $\eta_{\rm gal}$ varies
by less than a factor of $2$ over the factor of $\sim 30$ range in
virial mass spanned by the simulations. In the absence of feedback
$\eta_{\rm gal}$ peaks at low masses: feedback is clearly needed to
counter the high efficiency of gas cooling in low mass haloes.
We highlight however that the dependence of $\eta_{\rm gal}$ on
halo mass must become significantly stronger for halo masses below
those studied here (i.e. $M_{\rm vir} < 10^{11} h^{-1}M_\odot$) in order to
successfully reproduce the measured faint end of the luminosity function.

%%%%%%%%%%%%%%%%%%%%%%%%%%
\begin{center}
\begin{figure*}
\includegraphics[width=0.475\linewidth,clip]{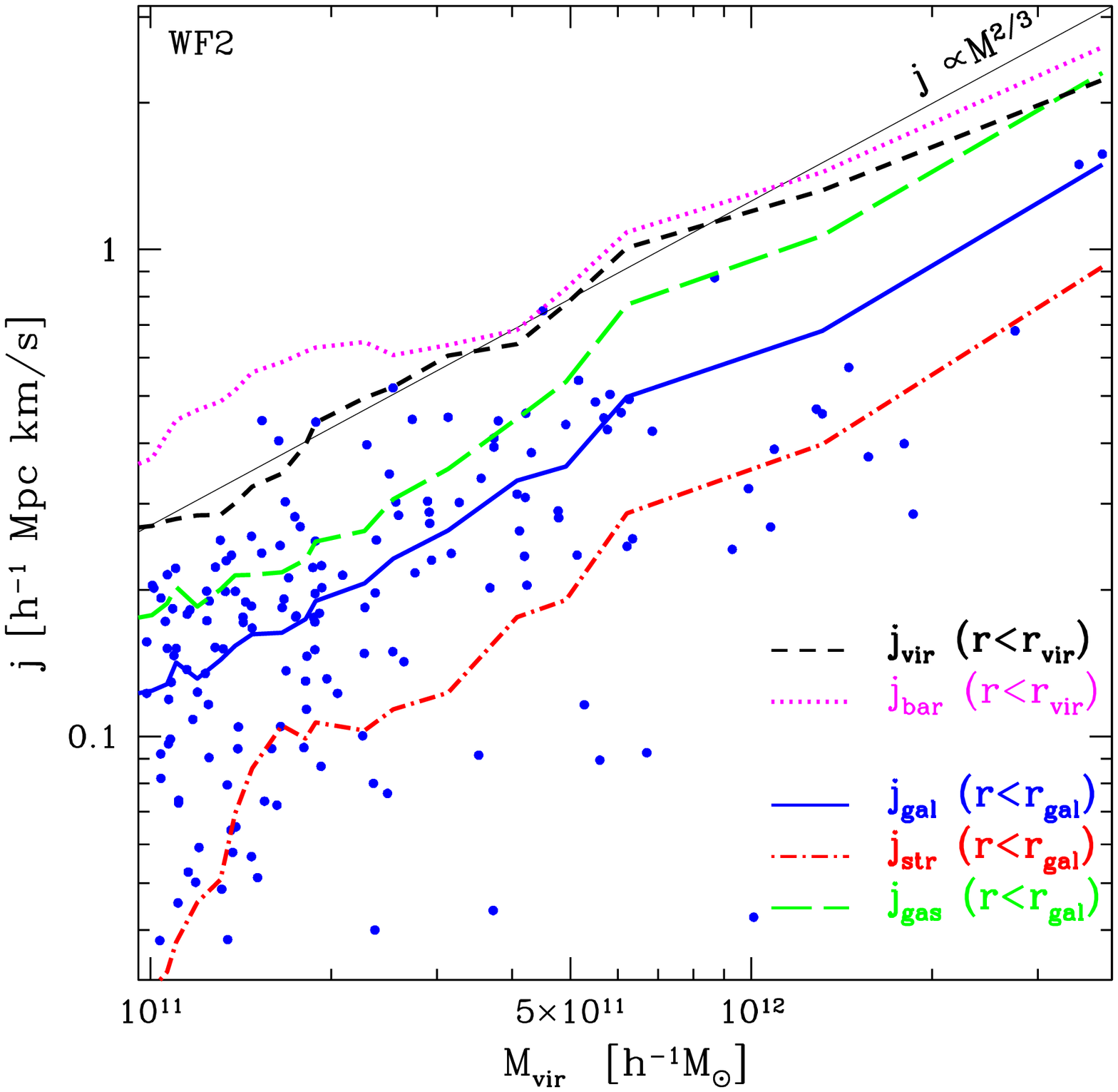}
\hspace{0.4cm}
\includegraphics[width=0.475\linewidth,clip]{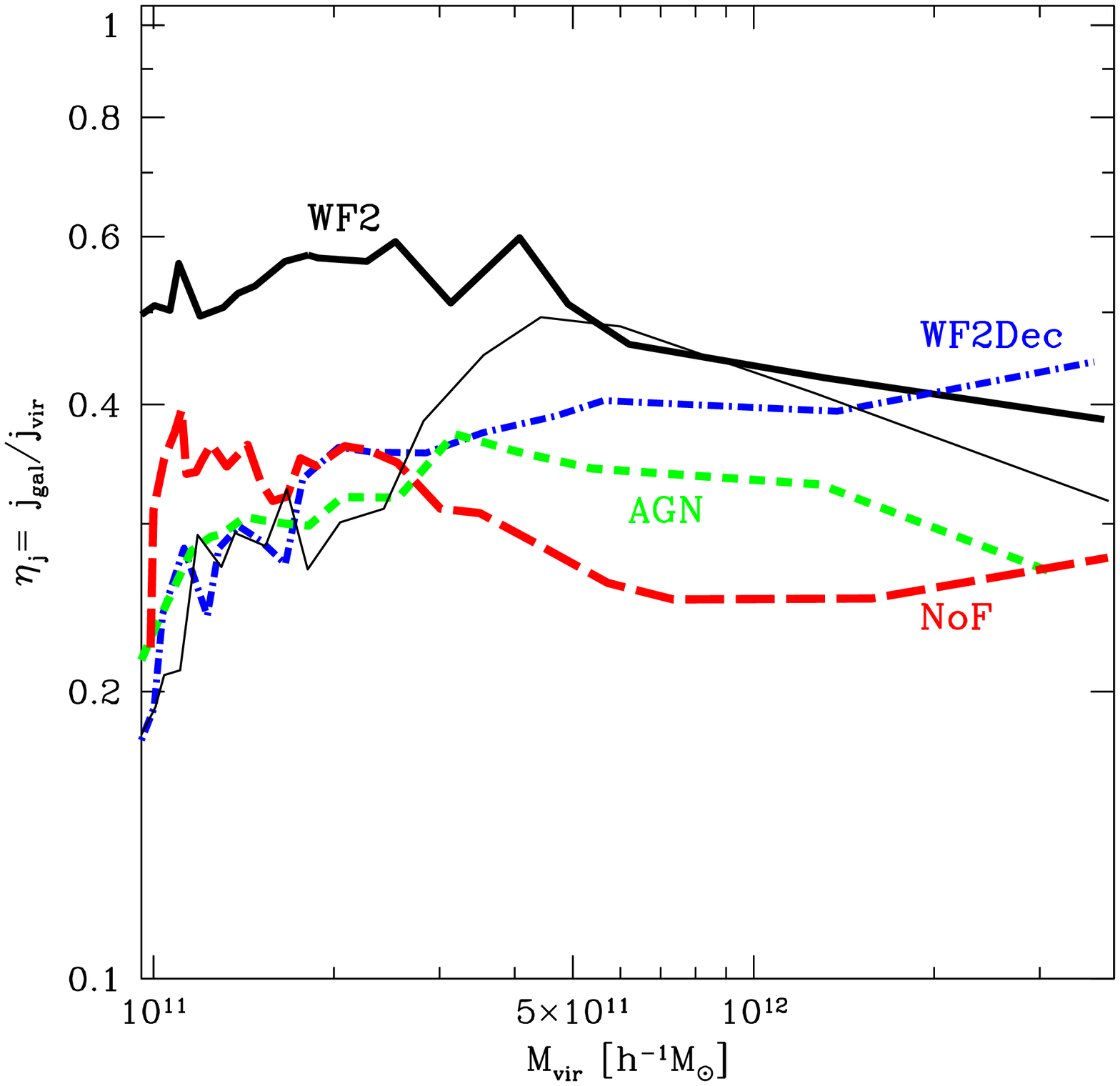}
\caption{ {\it Left:} Specific angular momentum, $j$, as a function of
  virial mass.  The black dashed line tracks the median $j$ of the
  dark matter component as a function of $M_{\rm vir}$. This follows
  closely the $j \propto M^{2/3}$ correlation expected for systems
  with constant spin parameter, $\lambda$. The other symbols, colors,
  and line types are the same as in Fig.~\ref{fig:BarM}. Note that the
  specific angular momentum of {\it all} baryons within $r_{\rm vir}$
  is quite similar to that of the halo as a whole (top dotted
  curve). The specific angular momentum of the central galaxy is
  typically lower than that of the halo; although it correlates well
  with $M_{\rm vir}$, the scatter is large.  {\it Right:} Angular
  momentum ``efficiency'', $\eta_j=j_{\rm gal}/j_{\rm vir}$, as a
  function of mass for various runs. For clarity, only the median of
  galaxies in runs NoF, WF2, WF2Dec, and AGN, are shown as a function
  of mass. Note that, unlike $\eta_{\rm gal}$, the angular momentum
  efficiency, $\eta_j$, is a weak function of both mass and of
  feedback. See text for further discussion.  }
\label{fig:BarJ}
\end{figure*}
\end{center}
%%%%%%%%%%%%%%%%%%%%%%%%%%

\subsection{Galaxy Angular Momentum}
\label{SecJGx}

The size and rotation speed of galaxy disks place powerful
observational constraints on galaxy formation models, and are directly
linked to the angular momentum acquired and retained by the baryons
that make up the galaxy. We explore this in Fig~\ref{fig:BarJ}, where
we show, in the left panel, the specific angular momentum of the
various galaxy components as a function of halo virial mass.

As in Fig.~\ref{fig:BarM}, dots correspond to the baryonic component
inside $r_{\rm gal}$ for individual systems in run WF2. Although the
scatter is large (an rms of $\sim 0.27$ dex), the solid curve, which
tracks the median $j$ as a function of $M_{\rm vir}$, shows that the
specific angular momentum scales roughly like $j \propto
M^{2/3}$. This is the same scaling found for the dark matter component
within $r_{\rm vir}$ (dashed black line), and is indeed the expected
scaling if the dimensionless halo spin parameter,
$\lambda=J|E|^{1/2}/GM^{5/2}$, is constant.

When all baryons within the virial radius are considered, their
specific angular momentum agrees well with that of the dark matter
(magenta dotted line). On the other hand, the specific angular
momentum of central galaxies is, on average, about $50\%$ that of its
surrounding halo. This fraction, which we refer to as the ``angular
momentum efficiency'', $\eta_j=j_{\rm gal}/j_{\rm vir}$, appears, on
average, to be roughly independent of halo mass for WF2 galaxies.

Interestingly, the gaseous and stellar components of WF2 galaxies have
distinctly different angular momenta. The gas has $2$ to $3$ times
larger specific angular momentum than the stars, implying that the
radial extent of gaseous disks in these galaxies is substantially
larger than that of the stellar component. This was already noted by
\citet{Sales2009} as a possible way to explain the large sizes of the
$z=2$ star-forming (gaseous) disks analyzed by
the SINS survey \citep{Forster2009}. We shall return to this issue in
Sec.~\ref{sec:obsdiag}.

\subsubsection{Feedback dependence}

The feedback dependence of the angular momentum efficiency, $\eta_j$,
is shown in the right panel of Fig.~\ref{fig:BarJ} as a function of
virial mass. Like the galaxy formation efficiency, $\eta_{\rm gal}$,
the halo mass dependence of $\eta_j$ is weak. Unlike $\eta_{\rm gal}$,
however, $\eta_j$ depends only weakly on feedback. In a given halo,
NoF galaxies have approximately the {\it same} angular momentum as
galaxies in the AGN run. This is striking, since their baryonic masses
differ on average by a factor of $\sim 10$. Feedback affects the
mass of a galaxy much more severely than its spin: 9 out of 10 baryons
in NoF galaxies are missing from AGN galaxies, but their specific
angular momenta are, on average, the same.

The thin line in the right panel of Fig.~\ref{fig:BarJ} shows the
results for WF2LR. Despite the large scatter, numerical resolution
effects are clearly noticeable below $\sim 4 \times 10^{11} h^{-1} \,
M_{\odot}$. This corresponds to $\sim 10^4$ particles per halo for WF2LR;
extrapolating this to WF2, it would mean that our results there should
be credible down to $\sim 5 \times 10^{10} h^{-1} \, M_{\odot}$. It
would therefore appear as if the main trends shown in
Fig.~\ref{fig:BarJ} are safe from resolution-induced numerical
artifacts.

The angular momentum efficiency peaks in moderate feedback runs (such
as WF2) at roughly $50\%$ and its dependence on feedback strength is
non-monotonic. Despite this apparent complexity, galaxy masses and
angular momenta are actually well correlated. Following
\citet{Sales2009}, we define the mass and angular momentum {\it
  fractions} $m_d$ and $j_d$, as 
\begin{equation}
m_d=\eta_{\rm gal} \, f_{\rm  bar}=M_{\rm gal}/M_{\rm vir},
\label{eq:md}
\end{equation}
and 
\begin{equation}
j_d={J_{\rm gal} \over J_{\rm vir}}={M_{\rm gal} \, j_{\rm gal} \over
  M_{\rm vir} \, j_{\rm vir}} = \eta_{\rm gal}\eta_{\rm j} f_{\rm bar}
\label{eq:jd}
\end{equation}
These parameters were introduced by \citet{Mo1998}, and have become
standard fare in semianalytic models of disk galaxy formation.

\citet{Sales2009} noted that $j_d$ and $m_d$ correlate well, but in a
manner different from the typical $j_d=m_d$ assumption of
semianalytic models \citep[e.g.,][]{Cole2000} and, perhaps more
importantly, insensitive to feedback.  These authors showed that the
simple expression 
\begin{equation}
j_d=9.71 \, m_d^2\, (1-\exp[-1/(9.71\,m_d)])
\label{eq:jdmd}
\end{equation}
provides a good approximation to the results of four {\small OWLS} runs with
supernova-driven winds: WF1, WF2, WF4 and WF2Dec.

\begin{center}
\begin{figure}
\includegraphics[width=84mm]{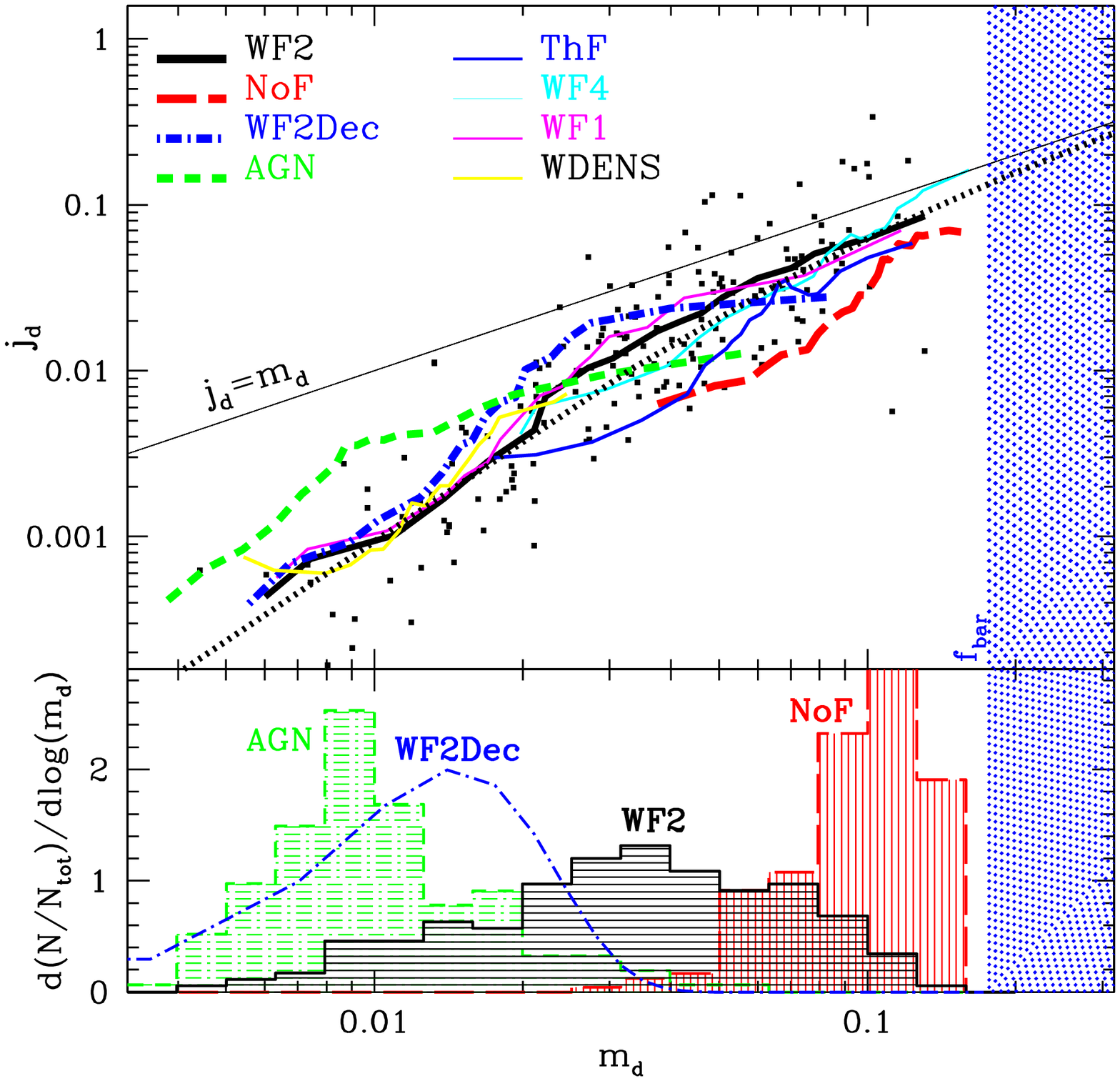}
\caption{ {\it Top:} The angular momentum fraction $j_d=J_{\rm
    gal}/J_{\rm vir}$ vs the galaxy mass fraction, $m_d=M_{\rm
    gal}/M_{\rm vir}$. Dots correspond to individual galaxies in WF2
  and are meant to illustrate the scatter;
  the median trend is traced by the black solid line. Other curves are
  analogous, but for each feedback model analyzed here. The black
  dotted curve is the fit proposed by \citet{Sales2009}. The straight
  line labeled $j_d=m_d$ corresponds to the commonly-adopted assumption
  that the specific angular momentum of a galaxy equals that of its
  surrounding halo. {\it Bottom:} Distribution of galaxy mass
  fraction, $m_d$, for four different runs spanning the range of
  feedback strengths of our simulations: NoF, WF2, WF2Dec, and AGN. }
\label{fig:JdMd}
\end{figure}
\end{center}

We revisit this result in Fig.~\ref{fig:JdMd}, where we show the
$j_d$-$m_d$ correlation for all the {\small OWLS} runs considered in this
paper. The dots show individual WF2 galaxies, and are meant to
illustrate the typical scatter in the relation; the curves trace the median
trend of $j_d$ with $m_d$ for the different runs while the black dotted curve outline
the relation in Eq.~\ref{eq:jdmd}. Although the AGN and NoF galaxies deviate
somewhat from the trend outlined in eq.~\ref{eq:jdmd} (indicated by
the dotted thick line), the departures are relatively small and the agreement
between runs seems remarkable given the extreme range in feedback models
explored here. 

The bottom panel in Fig.~\ref{fig:JdMd} shows the {\it distribution}
of $m_d$ for four different feedback implementations. Clearly,
feedback, at least as implemented in our models, affects mostly the
baryonic mass of galaxies, but largely preserves the link between the spins of
haloes and galaxies. This link imprints correlations between galaxy mass,
size, and rotation speed that may be contrasted with observations. We
turn to this issue next.

%%%%%%%%%%%%%%%%%%%%%%%%%%
%%%%%%%%%%%%%%%%%%%%%%%%%%

\section{Observational Diagnostics}
\label{sec:obsdiag}

\subsection{Size and Stellar Mass of $z=2$ Galaxies}
\label{ssec:rhmstr}

\begin{center}
\begin{figure*}
\includegraphics[width=0.475\linewidth,clip]{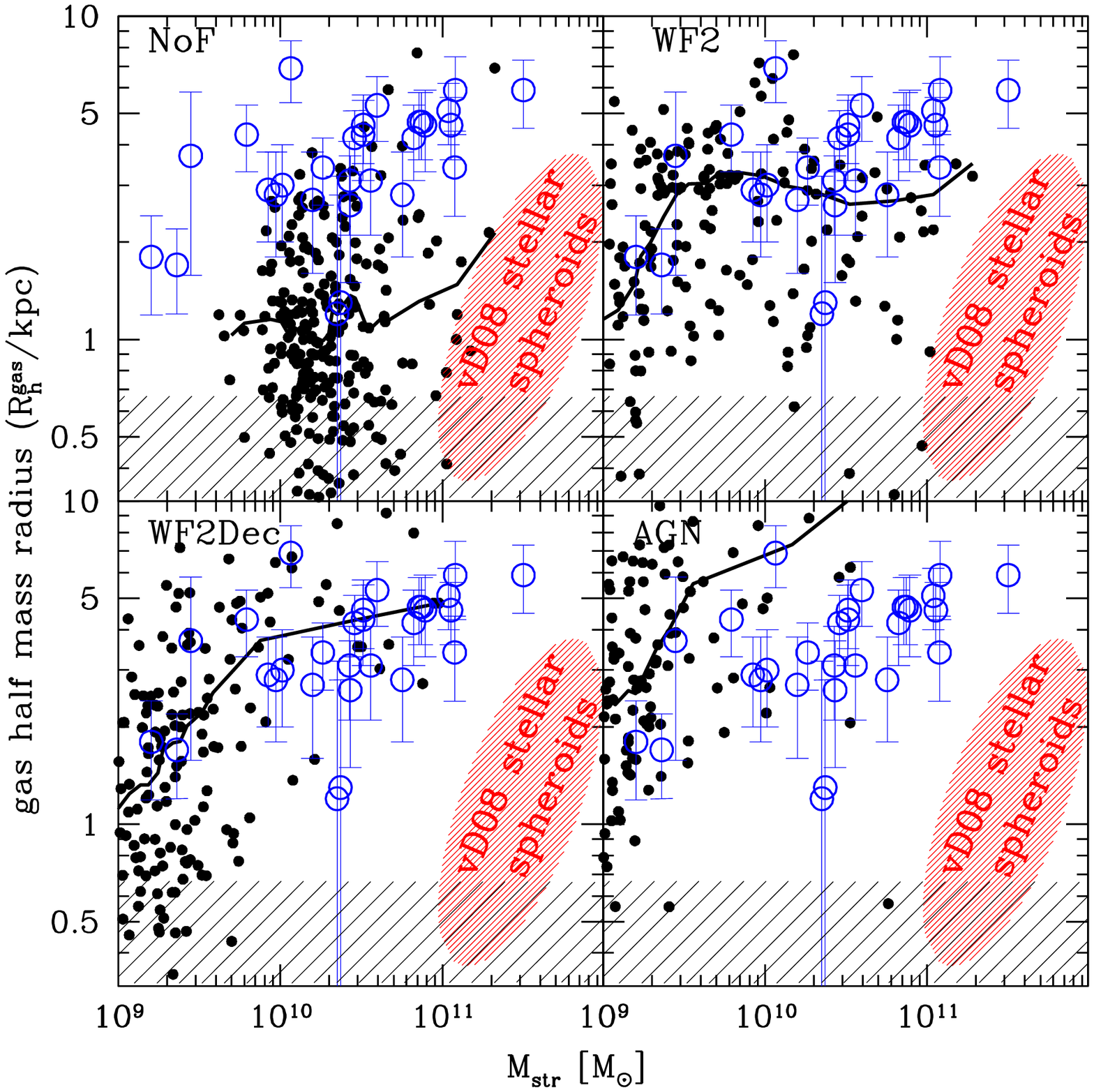}
\hspace{0.4cm}
\includegraphics[width=0.475\linewidth,clip]{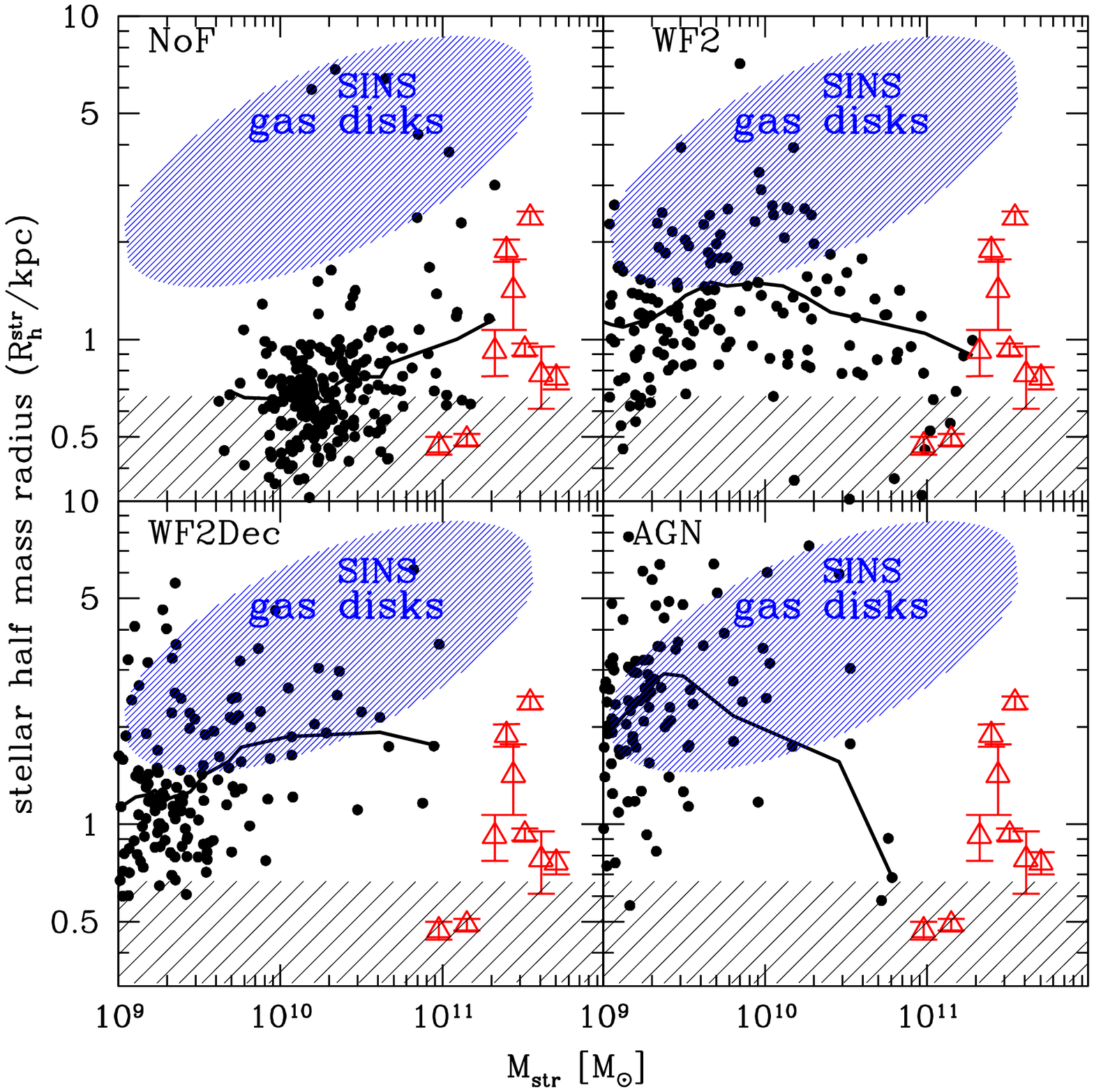}
\caption{ {\it Left:} half-mass radius of the gas as a function of
  stellar mass. Solid black dots in each panel show the results for
  four of our simulations NoF, WF2, WF2Dec and AGN. The thick solid
  line tracks the median as a function of mass. Open symbols with
  error bars correspond to the extended star-forming disk galaxies
  from the SINS survey \citep{Forster2009}, while the red shaded
  ellipsoid indicates the area of the plot occupied by the sizes of
  the compact quiescent galaxies from \citet{vanDokkum2008}. {\it
  Right:} same as before, but for the half-mass radii of the stars. In
  this case, open symbols and error bars are used to indicate the
  sizes of the compact stellar spheroids from \citet{vanDokkum2008},
  while the shaded blue area indicates the region of the plot occupied
  by extended gaseous disks from the SINS sample. The extended disks
  reported by SINS \citep{Forster2009} and the compact spheroidal
  galaxies from \citet{vanDokkum2008} probably bracket the size
  distribution of massive galaxies at $z=2$. The black shaded area
  indicates the gravitational softening of the simulations.  Half-mass
  radii for the gaseous components of simulated galaxies are typically
  larger than for the stars. Note that the size-stellar mass
  correlation is heavily dependent on feedback. When feedback is very
  efficient (e.g., WF2Dec/AGN) the size of the gaseous disks increase
  with stellar mass, a correlation that is reversed when feedback
  efficiency is low.}
\label{fig:RhMstr}
\end{figure*}
\end{center}

The feedback-driven trends of galaxy mass and angular momentum
efficiencies discussed above imprint different relations between
the stellar mass, $M_{\rm str}$, and the size of a galaxy. This is
shown in Fig.~\ref{fig:RhMstr}, where the various panels compare (for
runs NoF, WF2, WF2Dec, and AGN) the half-mass radius of the galaxy
versus $M_{\rm str}$. The panels on the left show the the half-mass
radius of the gas component whereas those on the right correspond to
the stars. The thick solid curve in each panel traces the median trend
as a function of $M_{\rm str}$. As noted above, simulated galaxies are
substantially more extended in gas than in stars.

The simulated galaxies are contrasted with data for the large
star-forming gas disks studied by the SINS survey
\citep[][hereafter FS09]{Forster2009}, as well as with the quiescent
compact red galaxies of \citet[][hereafter vD08]{vanDokkum2008}. These
two datasets probably bracket the extremes in the size distribution of
massive galaxies at $z=2$, from the most extended to the most compact.

When feedback is inefficient (e.g., the NoF run) most stars form in
dense, early-collapsing progenitors that merge later on to form the
final galaxies. During such mergers the baryonic component transfers
angular momentum to the surrounding halo, leading to the formation of
very compact massive galaxies
\citep{Navarro1991,Navarro1995,NavarroSteinmetz1997}. The galaxies
that result are therefore nearly as compact as the quiescent vD08
spheroids, although we note that many of those simulated galaxies have
half-mass radii even smaller than the gravitational softening of our
simulations, so their true sizes are actually uncertain. The gaseous
component in these simulations is also quite compact, with radii rarely
matching those of SINS disks.

Intermediate strength feedback (e.g., the WF2 run) has little effect on the most
massive galaxies, which are generally  as compact as the vD08 spheroids. On
the other hand, feedback affects more strongly less massive systems,
leading to a correlation between the mass and size of the stellar
component where, at the massive end, size decreases with increasing
mass. This trend runs counter the well-established galaxy scaling laws
at z=0 (brighter galaxies tend to be bigger). The trend is reversed at
lower masses and results, overall, in systems whose gaseous disks
overlap in properties with those of galaxies in the SINS survey.

Increasing the effects of feedback (as in the WF2Dec and AGN runs)
continues this trend, gradually reducing the mass of galaxies and
increasing their size at given $M_{\rm str}$. This is because the more
efficient the feedback the more massive the halo inhabited by a galaxy
of given stellar mass. More massive haloes are larger and have higher
specific angular momenta. Since, as we saw above, galaxies generally
inherit the specific angular momenta of their surrounding haloes, it is
possible to have fairly large galaxies of modest stellar mass because
they actually inhabit large, massive haloes. Indeed, many gaseous disks
in the AGN run are even more extended than the rather extreme examples
surveyed by SINS.

%%%%%%%%%%%%%%%%%%%%%%%%%%
%%%%%%%%%%%%%%%%%%%%%%%%%%

\subsection{The Tully-Fisher Relation at $z=2$}
\label{ssec:tf}

\begin{center}
\begin{figure}
\includegraphics[width=\linewidth,clip]{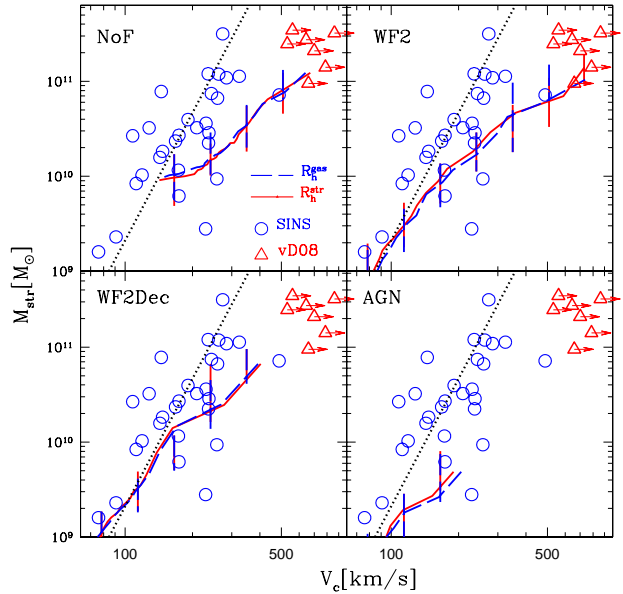}
\caption{ The stellar mass-circular velocity (Tully-Fisher) relation
  for $z=2$ galaxies identified in runs with four different feedback
  implementations. Symbols are as in Fig.~\ref{fig:RhMstr}. The median
  circular velocity measured at the stellar half-mass radius is shown
  by the solid red line. Vertical lines show the 25-75 percentiles of
  the distribution.  We also show this relation when the circular
  velocity is measured at the half-mass radius of the star forming gas
  (dashed blue curve).  Open circles and triangles show the
  observational determinations for disks and compact galaxies at z=2
  taken from \citet{Forster2009} and \citet{vanDokkum2008}.  For the
  latter we assign velocities by neglecting the dark matter
  distribution; i.e., we assume $V_c^2=G(M_{\rm str}/2)/R_{\rm eff}$,
  which constitutes a lower limit to the true circular velocity. This
  is indicated by the horizontal arrows in each panel. The thick
  dotted line is the \citet{Bell2001} relation for late-type galaxies
  at z=0 corrected to a Chabrier IMF.}
\label{fig:TF}
\end{figure}
\end{center}

The structural diversity of $z=2$ galaxies discussed above should also
be manifest in their kinematics. We explore this in Fig.~\ref{fig:TF},
where we plot, as a function of stellar mass, the circular velocity
estimated for SINS galaxies and for the compact vD08 galaxies. For
SINS, we use the ``maximum'' gas rotation speed, as quoted by FS09,
whereas for vD08 we estimate the circular velocity at the
effective radius based only on the contribution of the stellar
component; i.e., $V_c^2=G(M_{\rm str}/2)/R_{\rm eff}$. This is clearly
a {\it lower limit} to the circular velocity at that radius, since it
neglects the possible contributions of dark matter and gas
components. We note this in Fig.~\ref{fig:TF} by small arrows on the
vD08 data points (open triangles).

It is clear from this rendition of the data that the two populations
of $z=2$ galaxies follow very different Tully-Fisher relations. At
given stellar mass, the compact galaxies are expected to have circular
velocities at least {\it twice} higher than SINS disks. Although
kinematic data for such galaxies is scarce, \citet{vanDokkum2009}
report a preliminary measurement of the velocity dispersion of one of
these galaxies. The high velocity dispersion reported, $\sim 510$
km/s, agrees with this interpretation. 

The circular velocity of the simulated galaxies is measured at the
half-mass radius of the stellar (red solid curve) or the gaseous (blue
dashed curve) component, respectively. The comparison between
simulations and observations yields similar conclusions as in the
previous subsection.

Inefficient or absent feedback (e.g., NoF) yields galaxies that are
more concentrated than the SINS disks, and therefore have, at given
stellar mass, typically higher circular velocities.  Forming large,
extended disks is difficult in the absence of efficient feedback. By
contrast, accounting for the compact spheroids studied by vD08 is
relatively easy.

In the case of AGN or WF2Dec, the most efficient feedback schemes
explored in Fig.~\ref{fig:TF}, many simulated galaxies are as
spatially extended as the SINS disks, and the good agreement extends
to the Tully-Fisher relation for those galaxies. Few very massive
galaxies form as a result of the efficient feedback, and very few of
those that form are as compact as those in the vD08 sample.

More moderate feedback choices give intermediate results. We consider
it encouraging that some galaxies in the WF2 runs overlap with both
SINS and vD08 in Figs.~\ref{fig:RhMstr} and ~\ref{fig:TF}. If these
models are correct, then there should be a sizable population of
galaxies at $z=2$ with properties intermediate to the SINS disks and
vD08.

To summarize, the results shown in Figs.~\ref{fig:RhMstr} and
~\ref{fig:TF} indicate that neither the extreme compact sizes of
massive spheroids nor the large spatial extent of star-forming disks
at $z=2$ pose insurmountable challenges to the standard
paradigm. Indeed, it is possible, with adjustments to the feedback
algorithm, to reproduce either population without resorting to unusual
halo spin or halo formation histories.

At the same time, reproducing the striking diversity in the observed
sizes and masses of $z=2$ galaxies with a single feedback recipe might be
challenging, but we are encouraged by the large scatter in the
properties of simulated galaxies at given stellar mass that arises
naturally in {\it any} feedback model. The relative
abundance of either population is still poorly constrained
observationally, and our small simulation box might not be adequate to
study or search for rare, extreme populations. Improved observational
constraints on the relative abundance of extended vs compact galaxies
and a better characterization of the ``average'' population of $z=2$
galaxies will certainly help to constrain which feedback
implementation gives results that agree best with observation.

%%%%%%%%%%%%%%%%%%
%%%%%%%%%%%%%%%%%%%%%%%%%%

\subsection{Disks and Mergers at $z=2$}
\label{ssec:dskmrg}

Another interesting constraint is provided by the frequency of systems
actively forming stars in rotationally-supported disks. Before
surveys such as SINS and OSIRIS \citep[][ and references
therein]{Forster2009,Law2009} started to resolve the kinematics of
star-forming galaxies at high $z$, it had been commonplace to assume
that systems where star formation was progressing in earnest would
almost invariably be ongoing major mergers. It is now clear, however,
that at least about one third of the galaxies surveyed by SINS and
OSIRIS are forming stars in relatively quiescent disks
rather than ongoing mergers with disturbed and transient kinematics
\citep[for an alternative view, however,
see][]{Robertson2008}.

\begin{center}
\begin{figure*}
\includegraphics[width=0.2375\linewidth,clip]{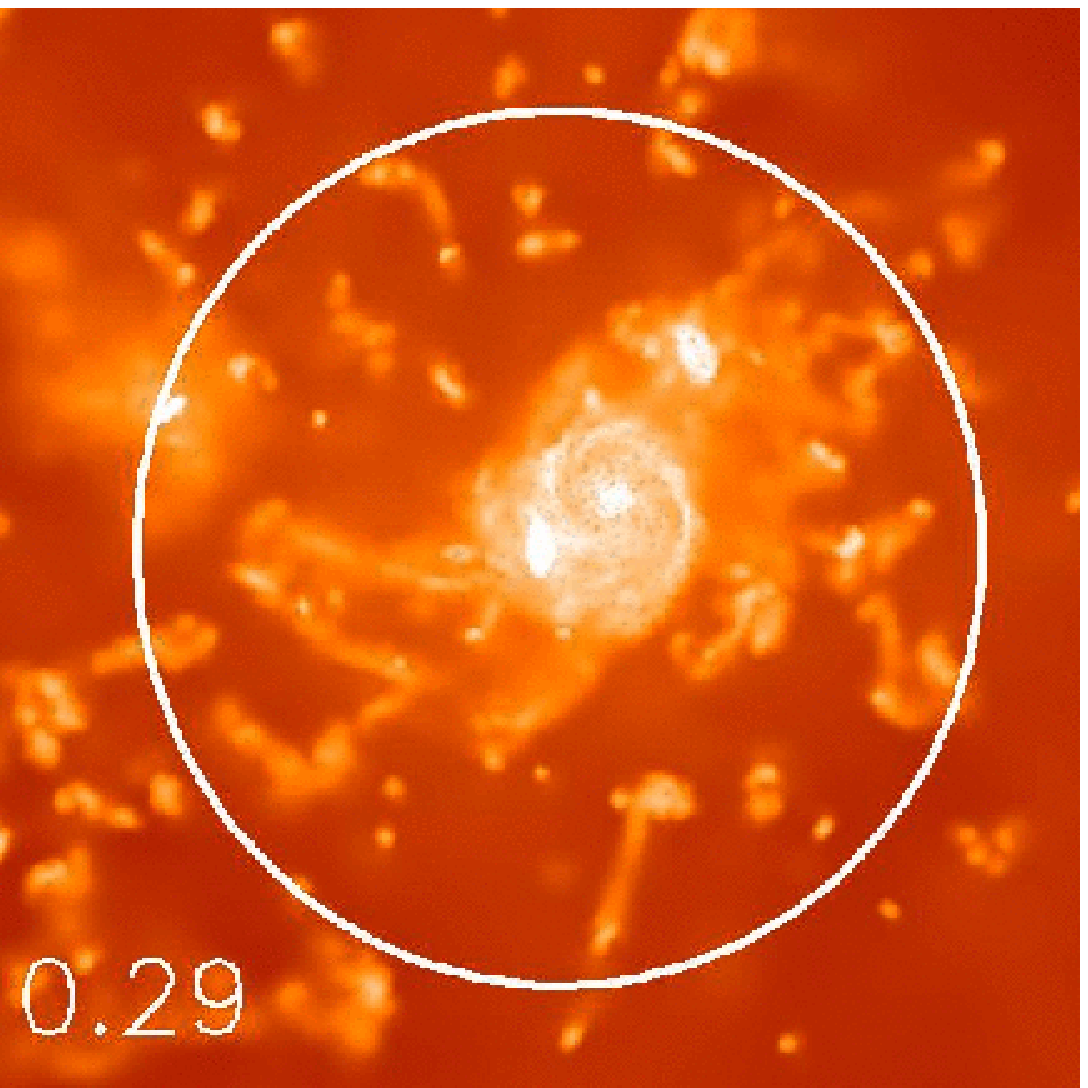}
\includegraphics[width=0.2375\linewidth,clip]{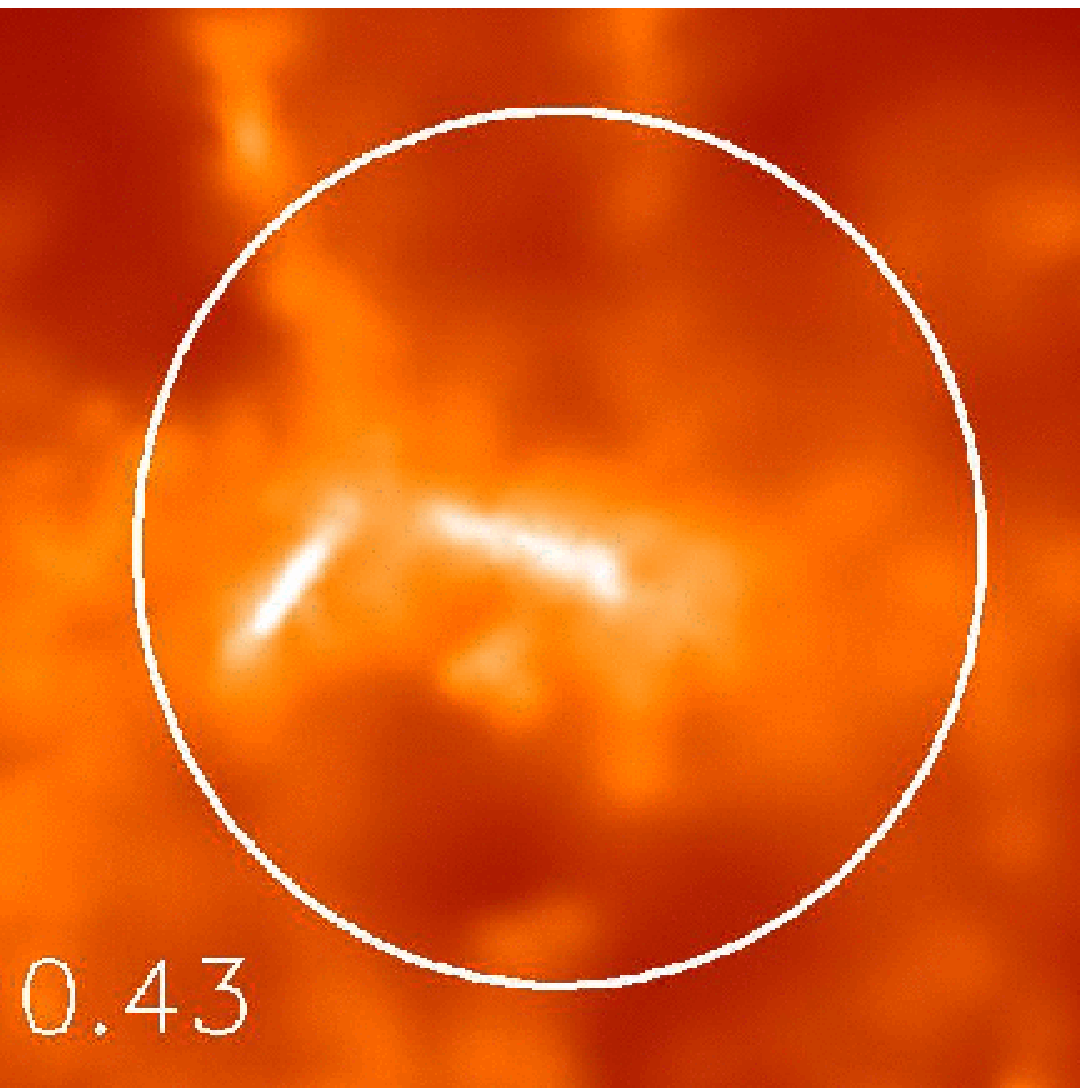}
\includegraphics[width=0.2375\linewidth,clip]{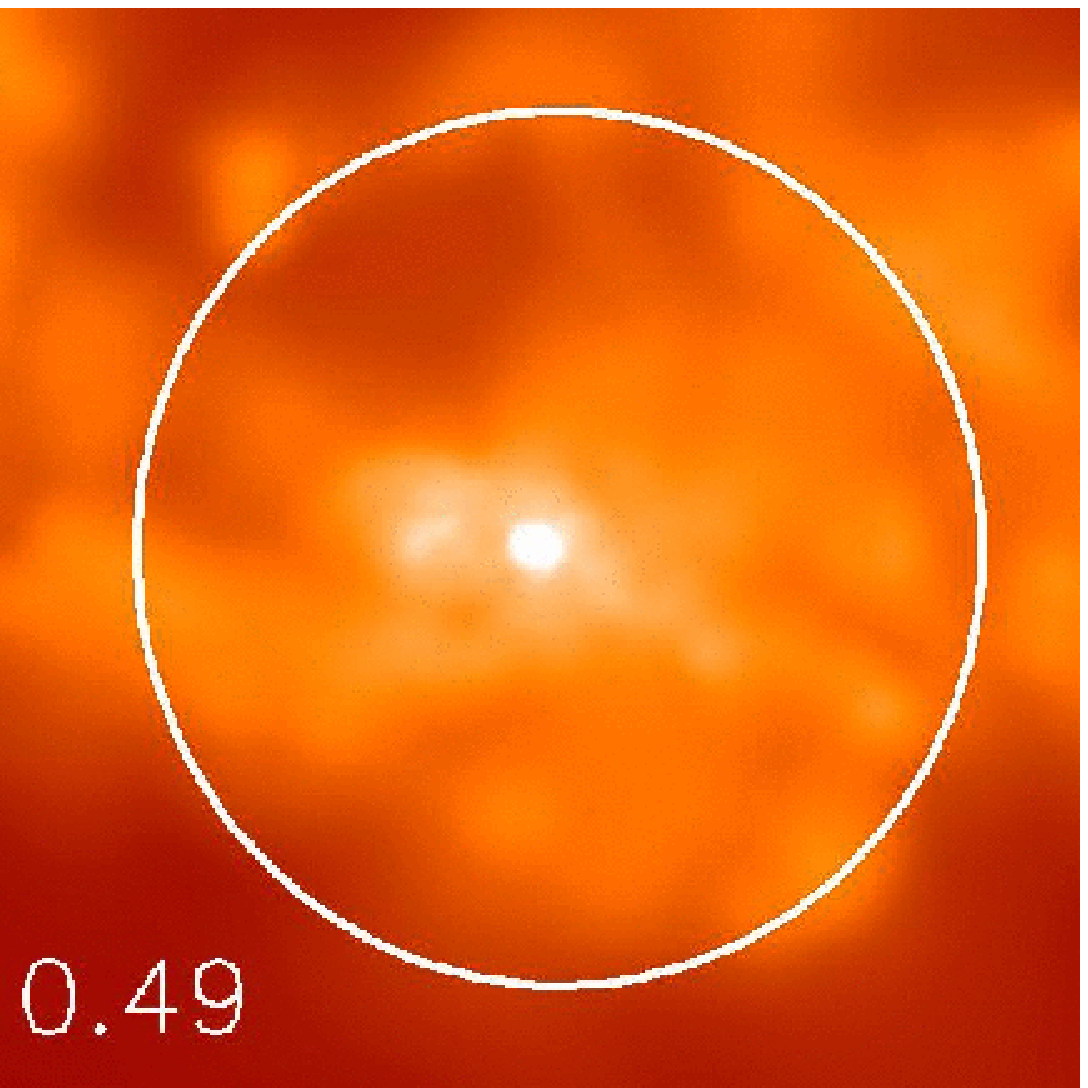}
\includegraphics[width=0.2375\linewidth,clip]{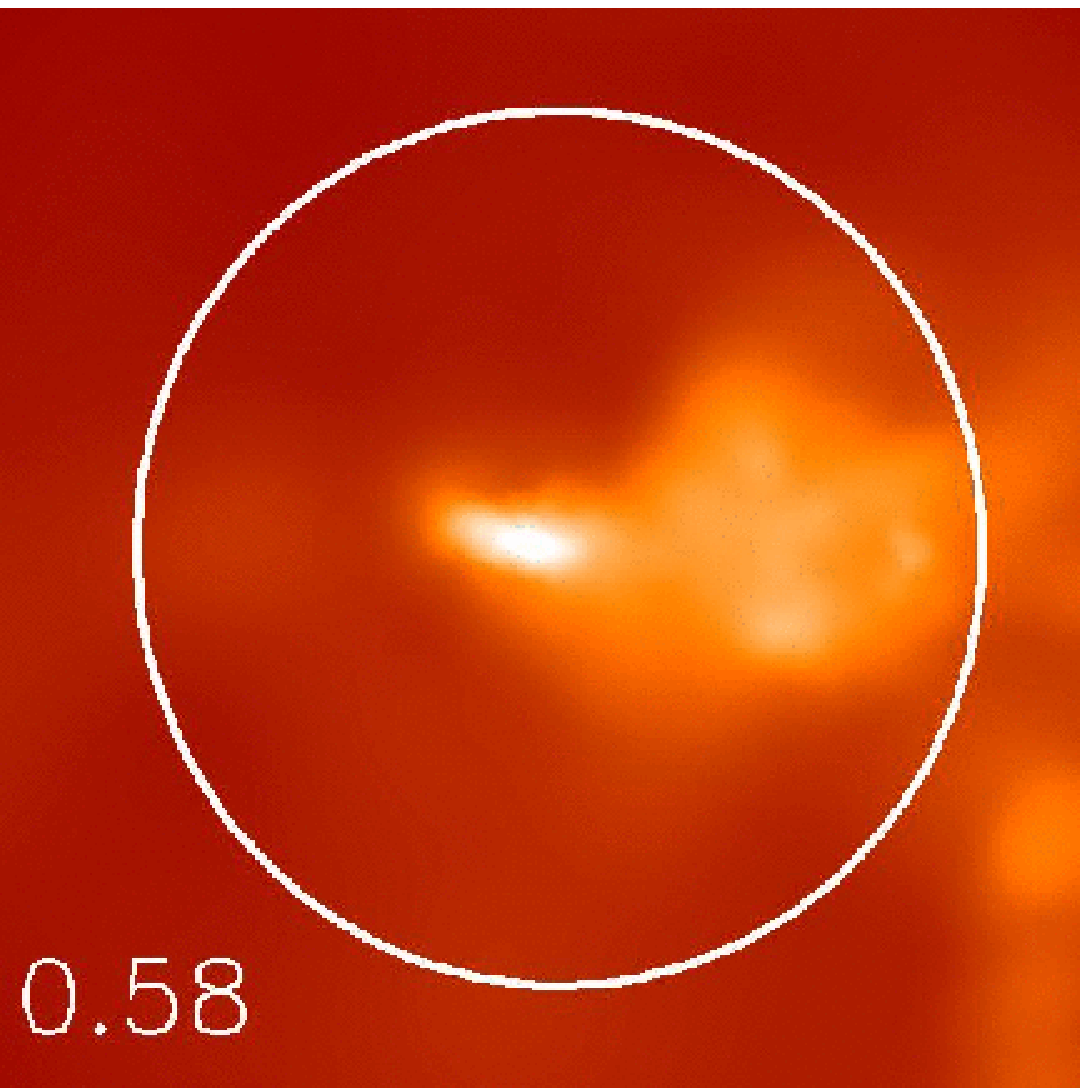}
%% second row
\includegraphics[width=0.2375\linewidth,clip]{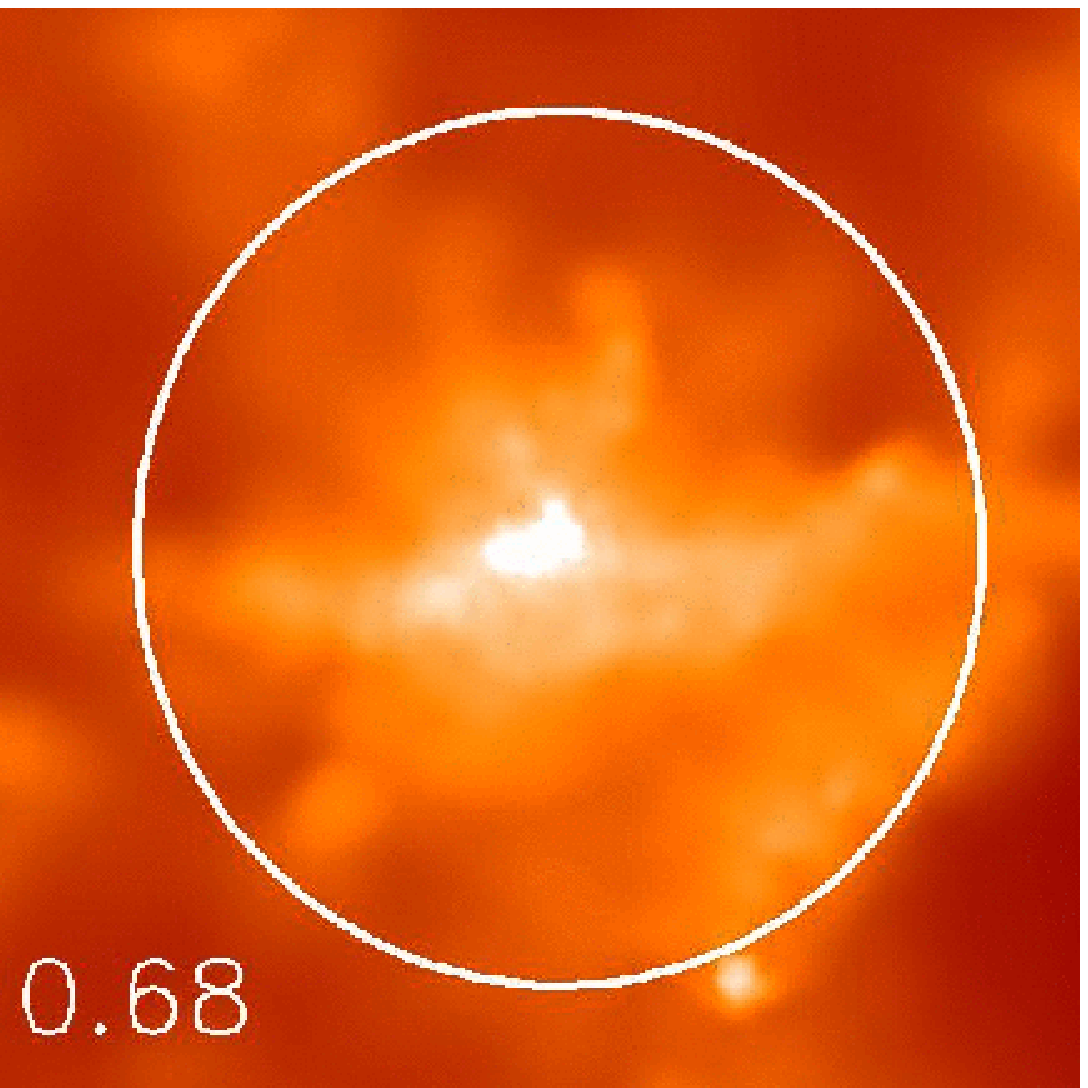}
\includegraphics[width=0.2375\linewidth,clip]{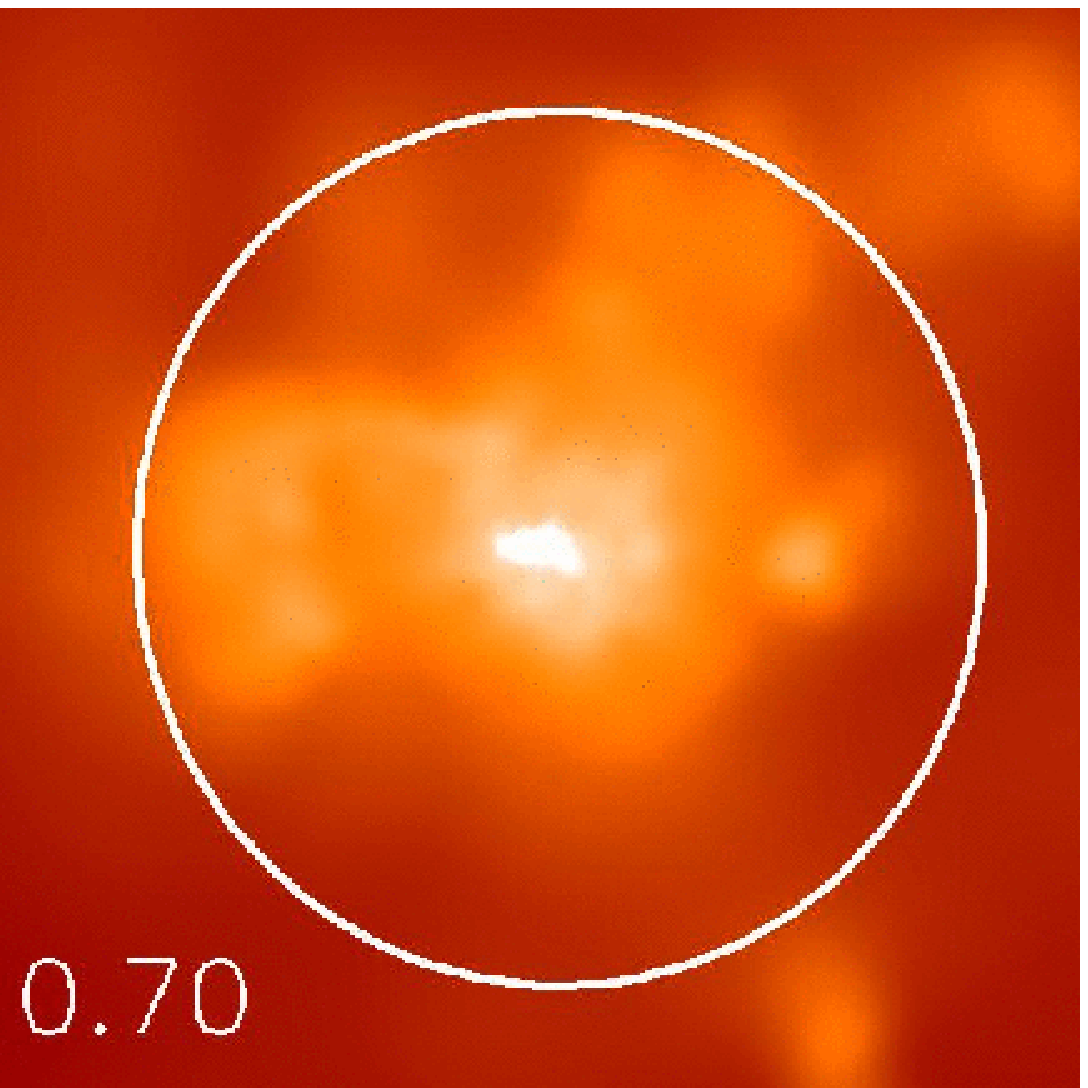}
\includegraphics[width=0.2375\linewidth,clip]{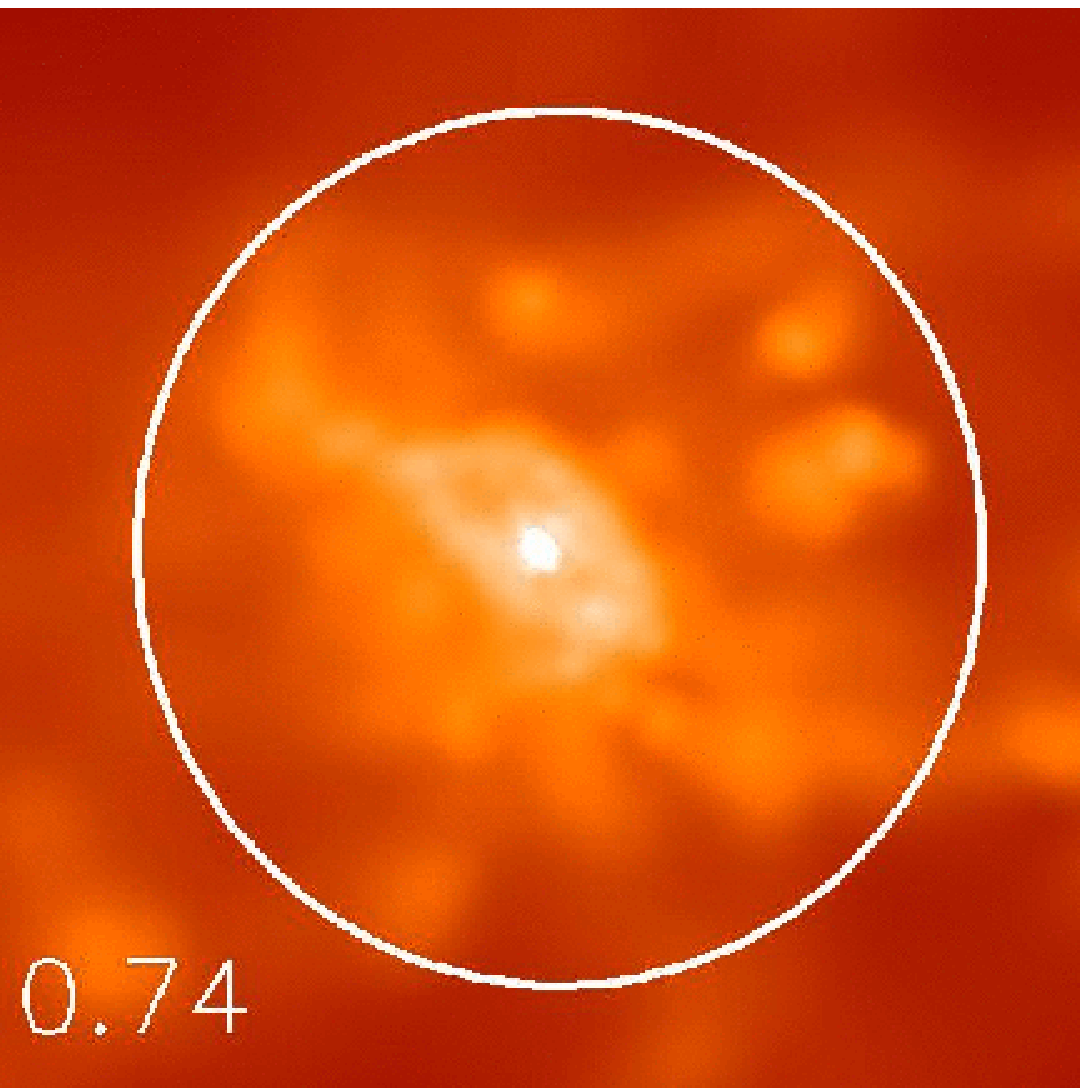}
\includegraphics[width=0.2375\linewidth,clip]{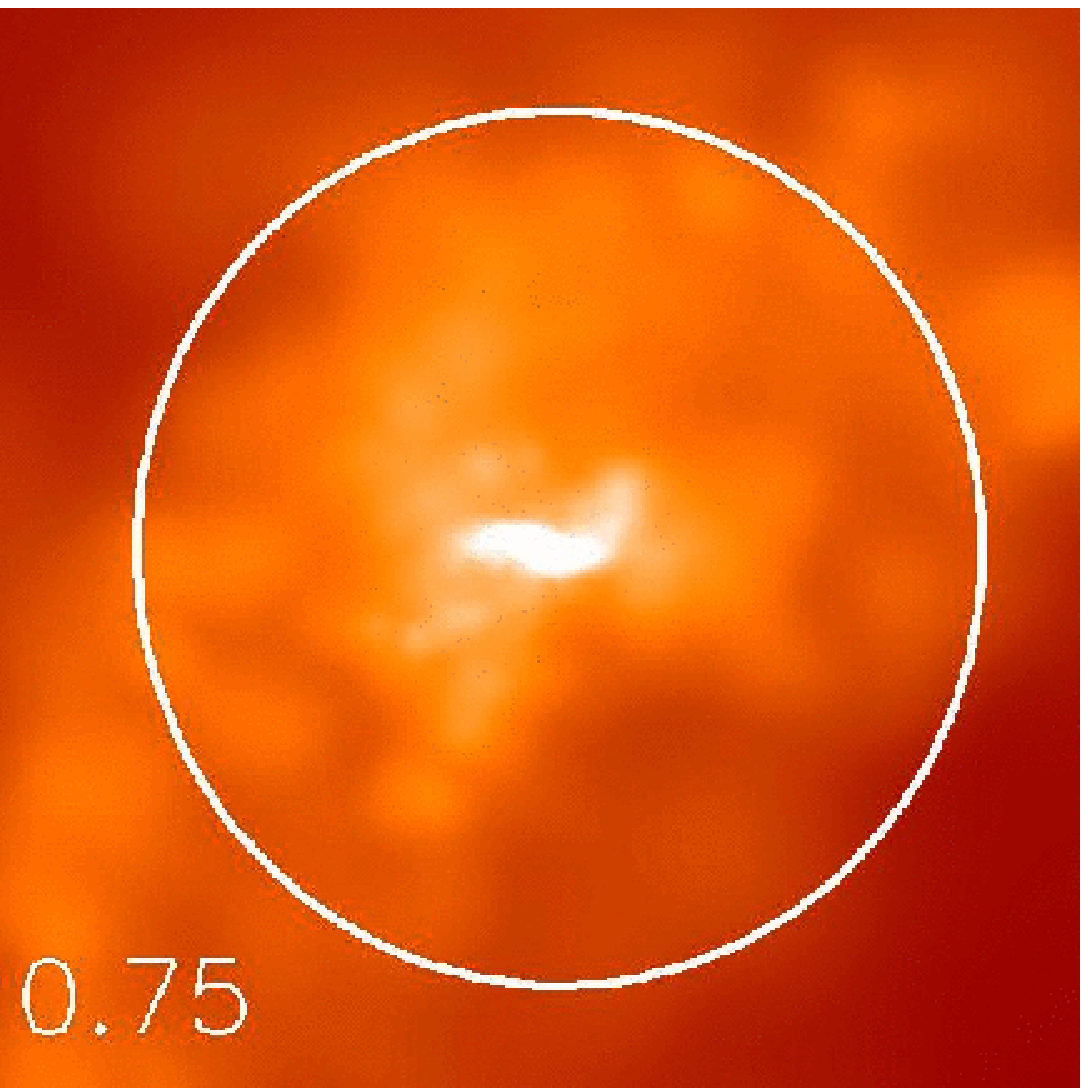}
% third row
\includegraphics[width=0.2375\linewidth,clip]{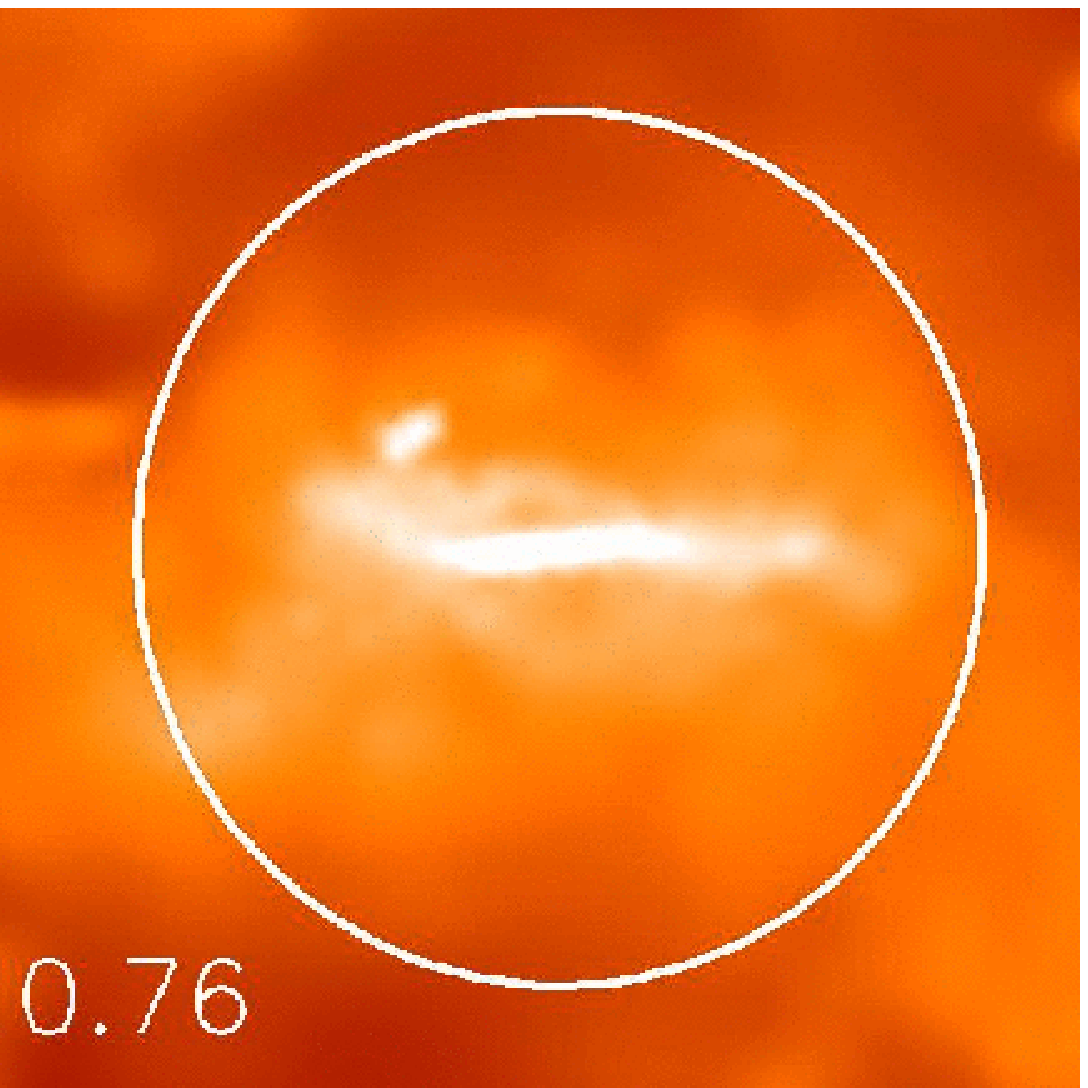}
\includegraphics[width=0.2375\linewidth,clip]{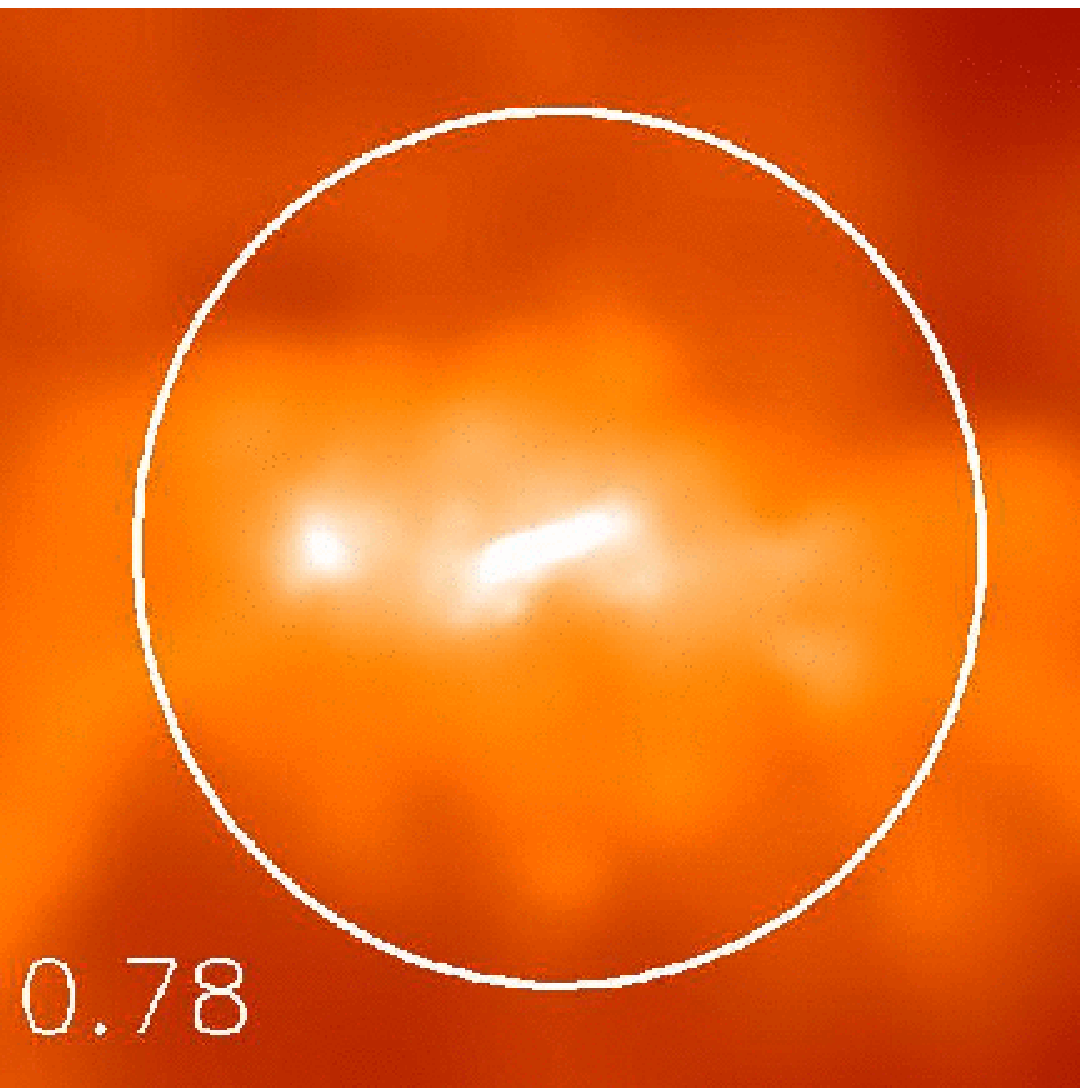}
\includegraphics[width=0.2375\linewidth,clip]{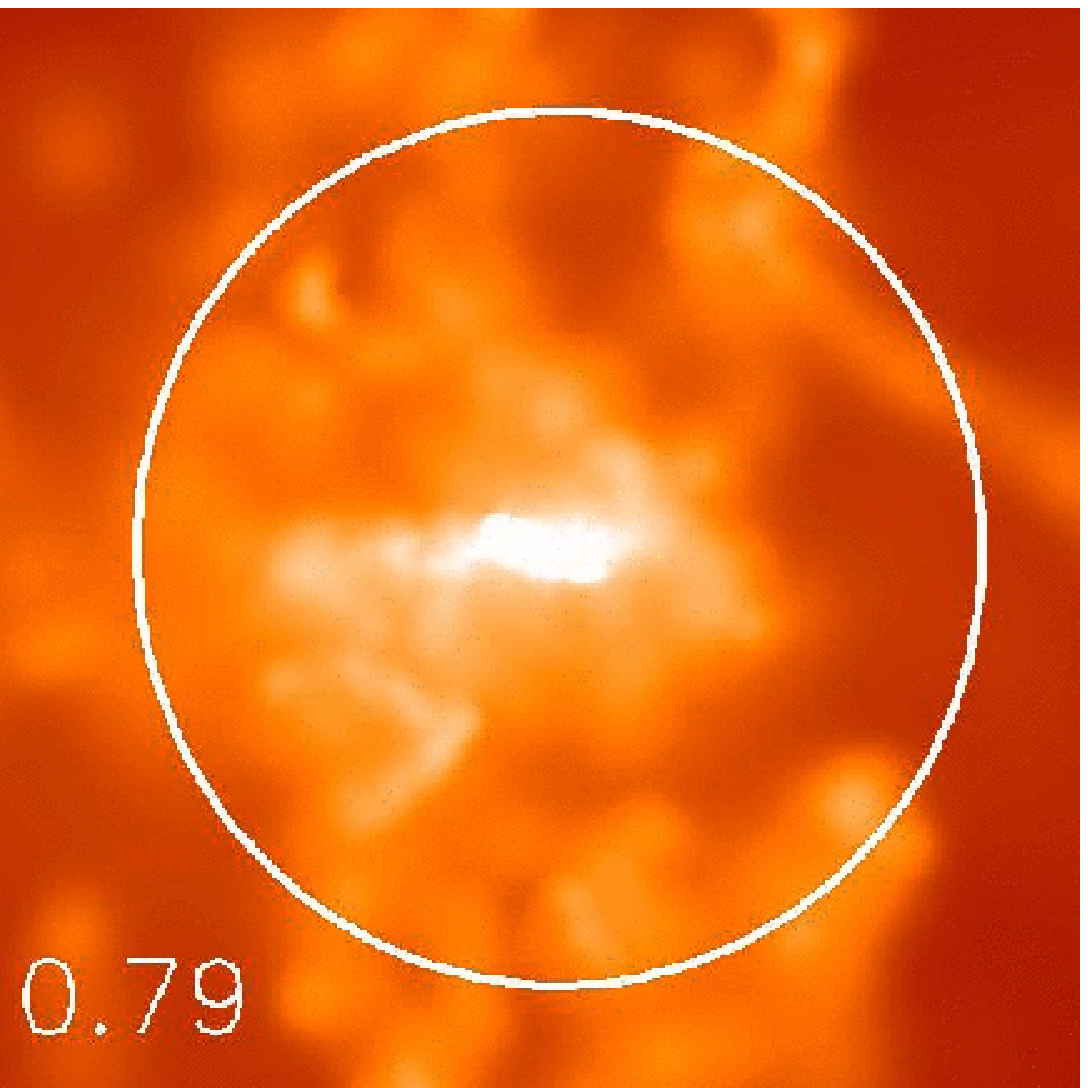}
\includegraphics[width=0.2375\linewidth,clip]{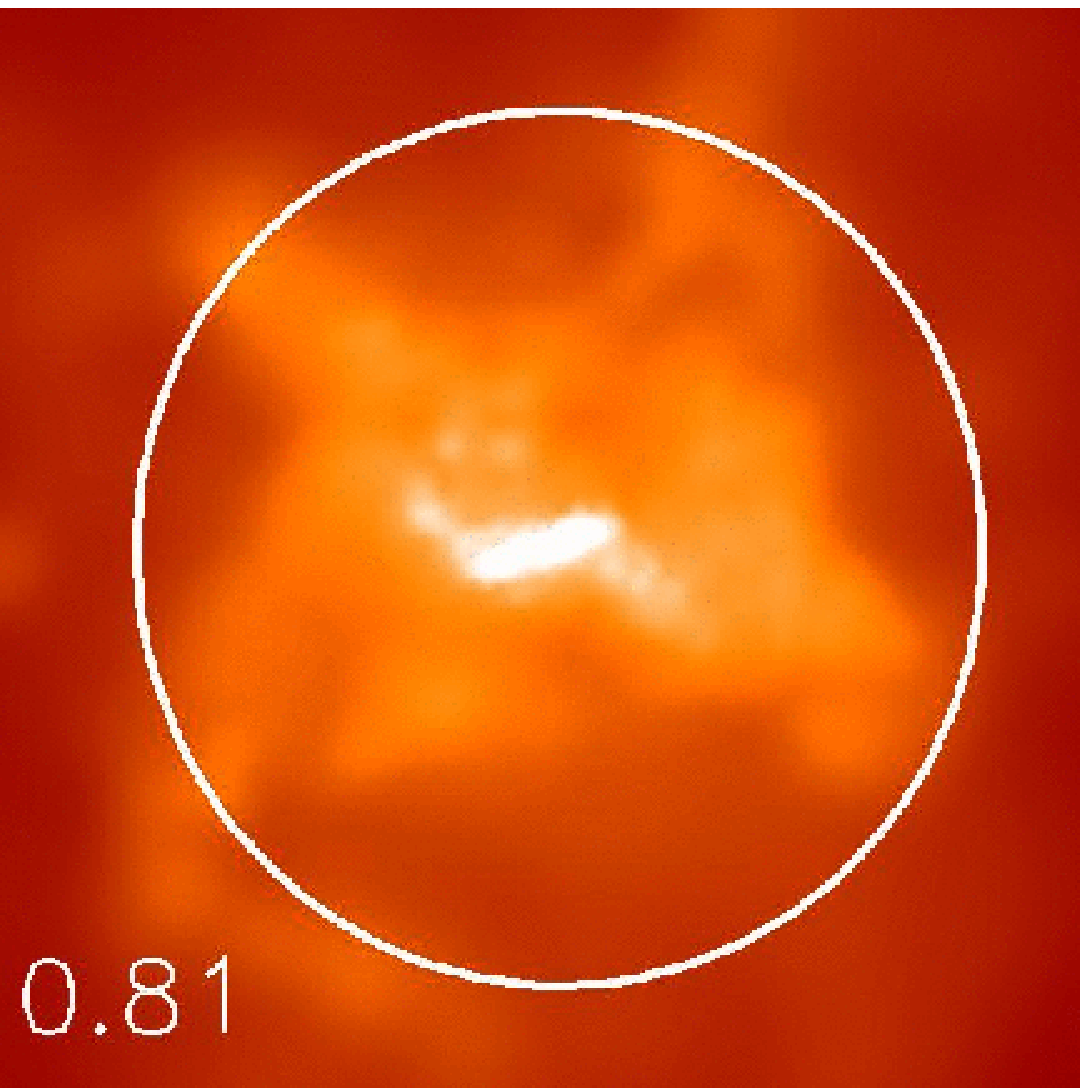}
% fourth row
\includegraphics[width=0.2375\linewidth,clip]{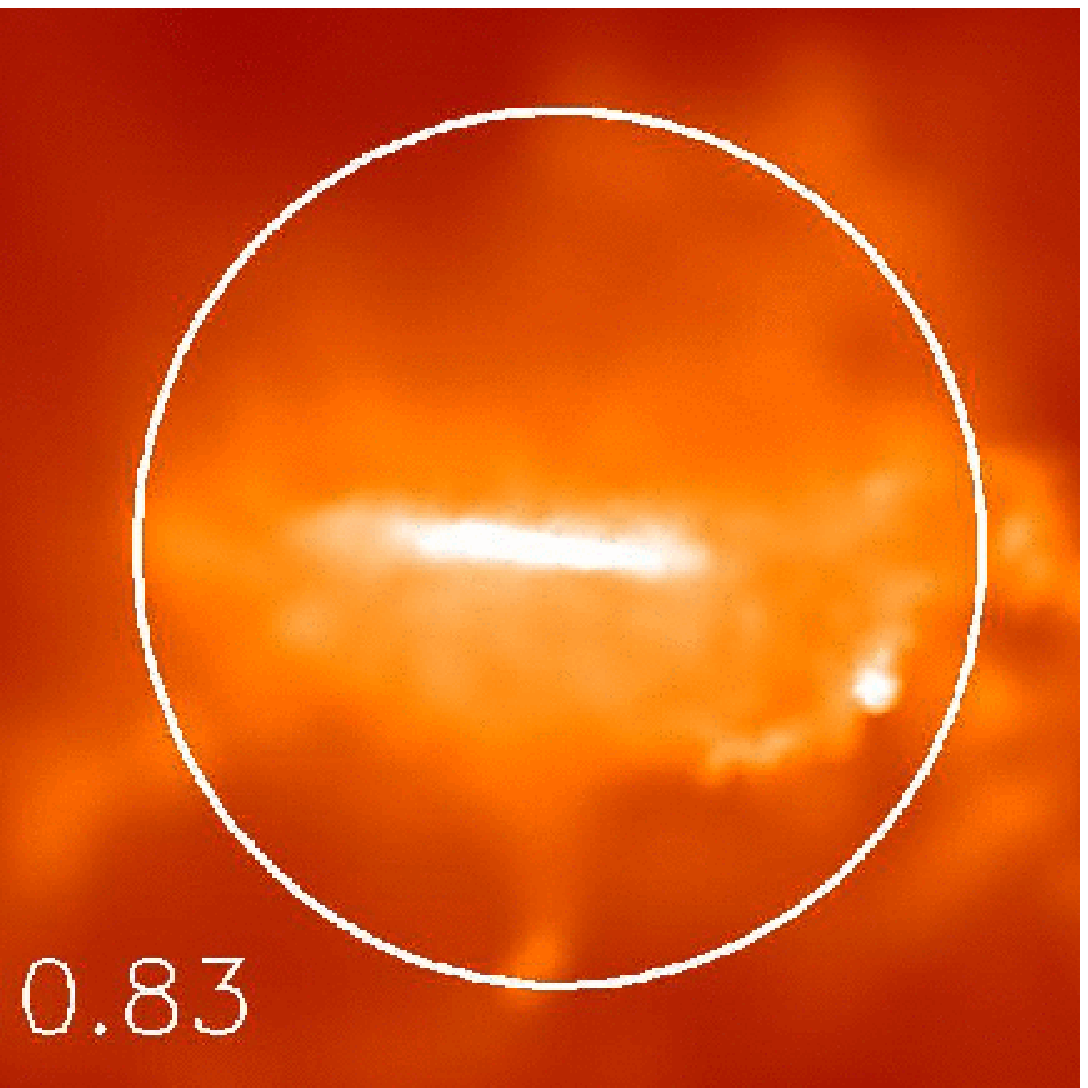}
\includegraphics[width=0.2375\linewidth,clip]{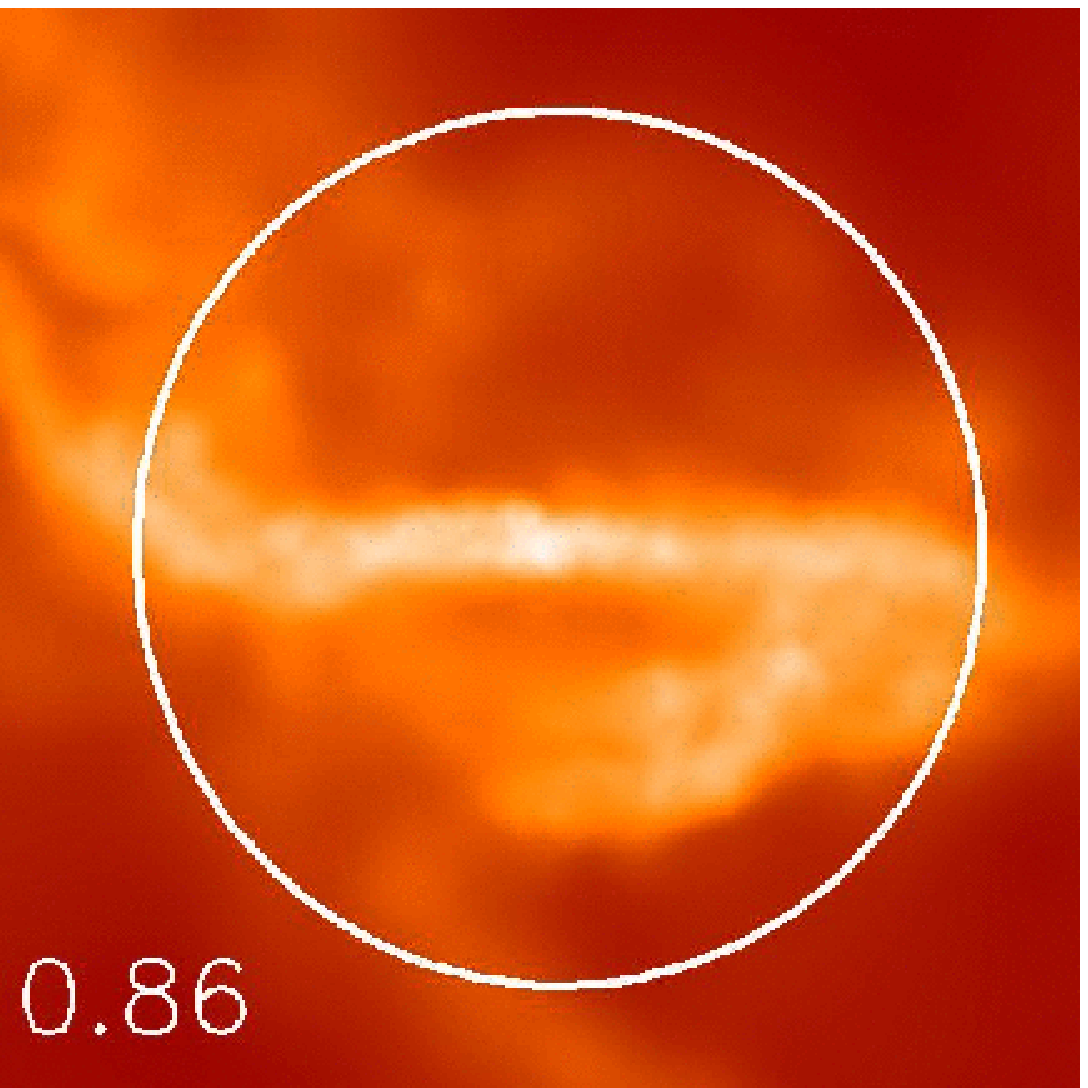}
\includegraphics[width=0.2375\linewidth,clip]{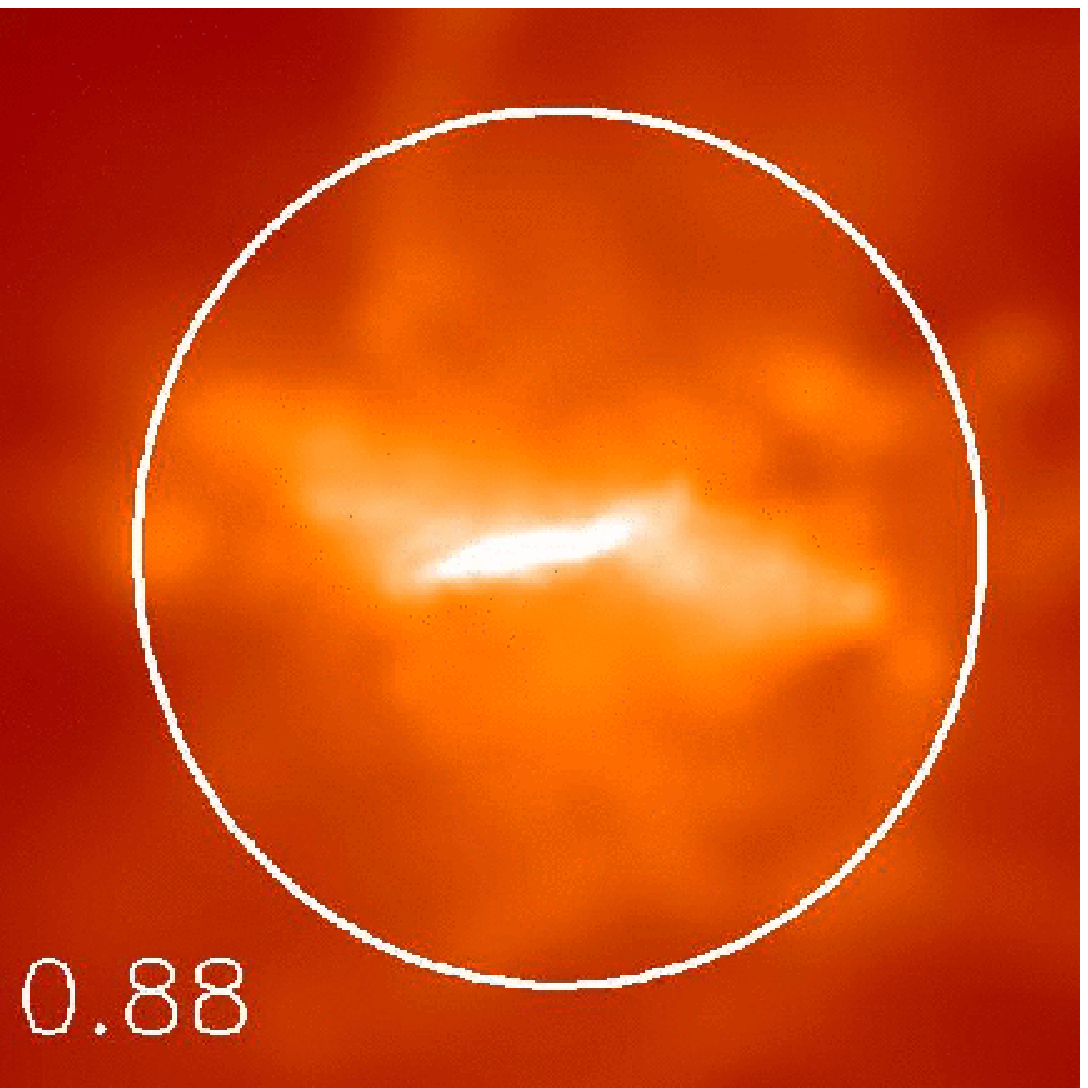}
\includegraphics[width=0.2375\linewidth,clip]{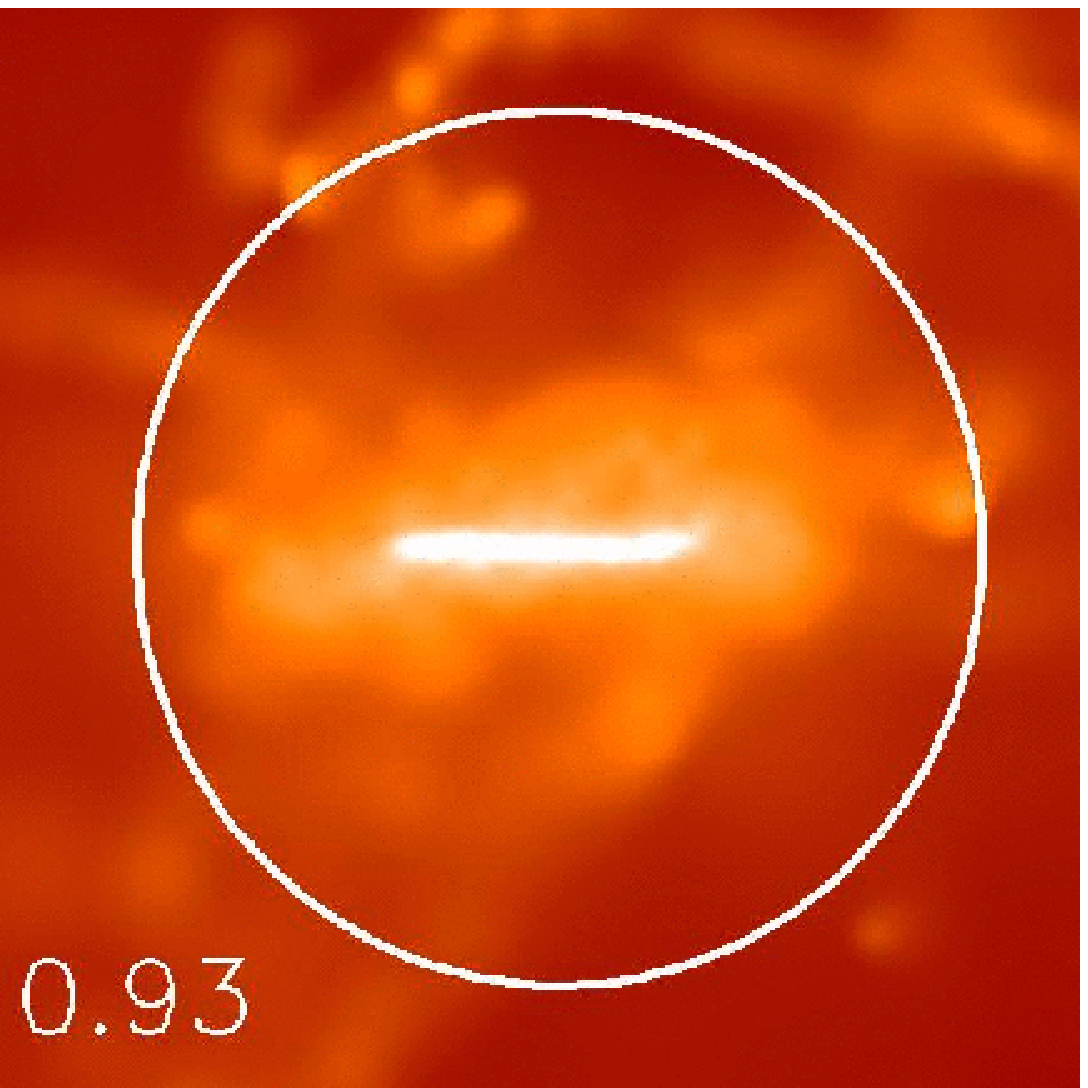}

\caption{ Edge-on view of galaxies spanning a wide range in rotational
  support taken from the WF2 run. Each panel is labelled by the value of $\kappa_{\rm rot}$
  of the star-forming gas. Colors are assigned according to the (projected)
  logarithmic densities of the gas. Well-defined disk systems are
  apparent when $\kappa_{\rm rot}\gsim 0.75$; lower values of this
  parameter indicate ongoing mergers and/or systems with disturbed
  morphology. Solid circles indicate the region selected as the galaxy radii:
  $r_{\rm gal} = 0.15 \, r_{\rm vir}$.}
\label{fig:RotMorph}
\end{figure*}
\end{center}

We use $\kappa_{\rm rot}$, the simple measure of the importance of
ordered rotation introduced in Sec.~\ref{SecFbMorph}, to explore this
issue in our simulations. When most of the gas is in a
rotationally-supported disk, the parameter $\kappa_{\rm rot}$ should
approach unity. Fig.~\ref{fig:RotMorph} enables a visual calibration
of this parameter by showing edge-on projections of 12 galaxies
arranged by the value of $\kappa_{\rm rot}$ of the central galaxy (in
this case only the star-forming gas is used to compute $\kappa_{\rm
rot}$).  Figure~\ref{fig:RotMorph} shows an image of the projected gas
density within a sphere of radius $1.3 \, r_{\rm gal}$. Thin, extended
disks are the norm when $\kappa_{\rm rot} \gsim 0.75$. Ongoing mergers
typically have $\kappa_{\rm rot} \lsim 0.5$; those with intermediate
values of $\kappa_{\rm rot}$ have disturbed morphologies, and tend to
be late-stage mergers or systems where accretion is ongoing but minor.

\begin{center}
\begin{figure}
\includegraphics[width=84mm]{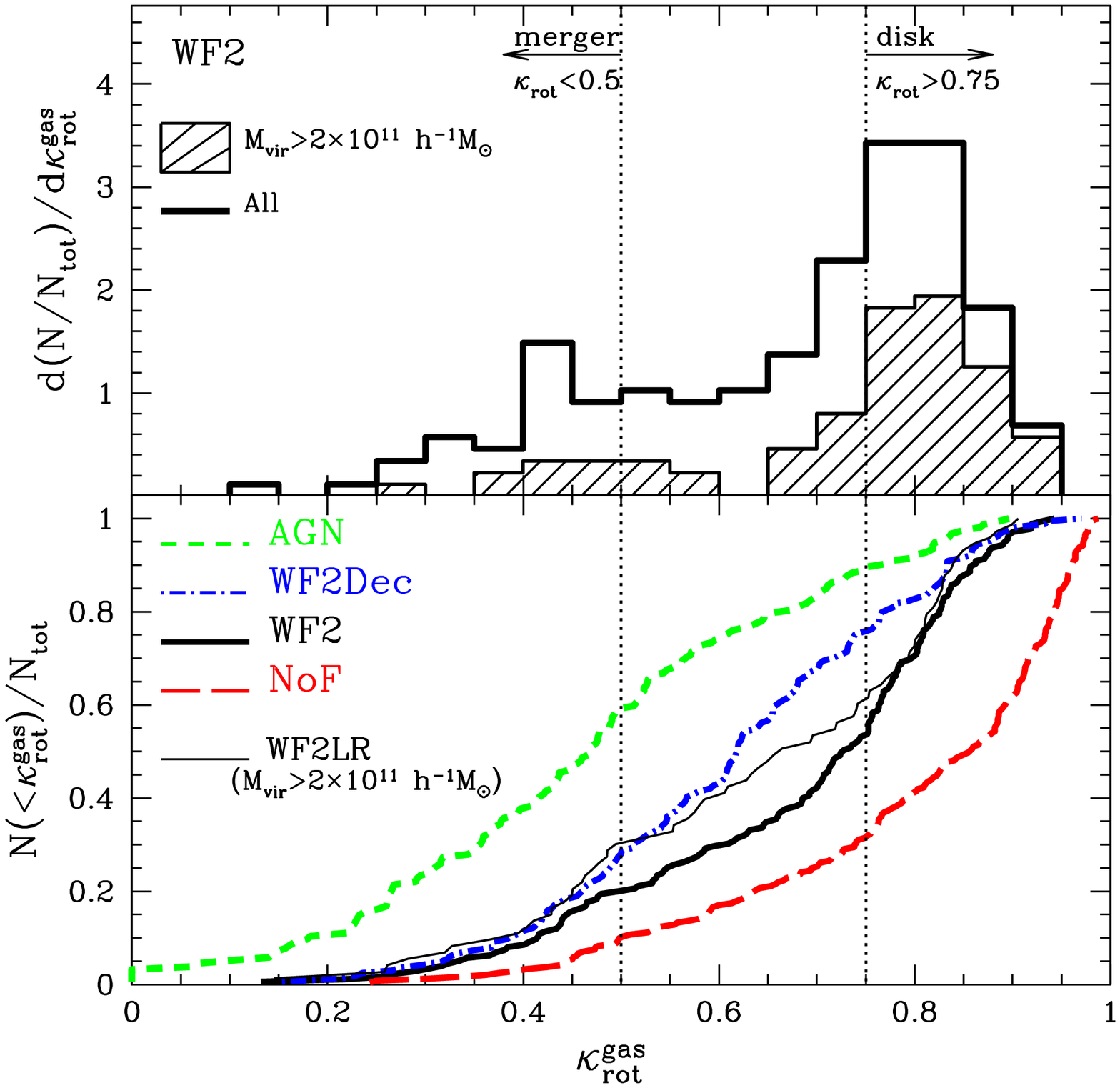}
\caption{ Upper panel shows the histogram of $\kappa_{\rm rot}=K_{\rm rot}/K$, the fraction
  of kinetic energy of star-forming gas particles in ordered rotation
  for all galaxies in the WF2 run. The shaded red histogram is the same,
  but only for the half most massive, and therefore best numerically
  resolved, systems. The similarity between the two suggests that
  numerical resolution does not play a significant role in the
  statistics. $\kappa_{\rm rot}$ should be approximately unity for a
  disk where all particles are in circular orbits and much smaller for
  systems where ordered rotation plays a less important role. The
  large number of systems around $\kappa_{\rm rot} \sim 0.8$ indicates
  that systems where star formation occurs in well-defined disks are
  quite common in this run (see Fig.~\ref{fig:RotMorph} for
  examples). The cumulative fraction of systems as a fraction of
  $\kappa_{\rm rot}$ for the four different feedback implementations 
  are shown in the bottom panel. A trend for gaseous disks becoming more
  prevalent as feedback efficiency decreases is clearly seen.}
\label{fig:HistErot}
\end{figure}
\end{center}

Using this simple measure, the fraction of ongoing mergers vs
quiescent disks may be readily estimated, and is shown in the top
panel of Fig.~\ref{fig:HistErot} for the case of WF2. The
distribution of $\kappa_{\rm rot}$ for all WF2 galaxies is shown by
the top histogram; the shaded histogram is for the same run, but
reducing the sample of galaxies to one half by selecting only those in
haloes more massive than $2 \times 10^{11} h^{-1}
M_{\odot}$. Encouragingly, the shape of the two histograms is quite
similar. This is further confirmed by the distribution of $\kappa_{\rm
  rot}$ in WF2LR galaxies (for $M_{\rm vir}>2\times 10^{11} h^{-1} \,
M_{\odot}$) which is shown as the thin solid line in the bottom panel of
Fig.~\ref{fig:HistErot} . The good agreement between WF2 and WF2LR
indicates that numerical resolution effects are unlikely to compromise
our conclusions.

According to the definition above, about $45\%$ of WF2 galaxies are
reasonably quiescent star-forming disks, and only about $20\%$ are
ongoing major mergers. These fractions are similar for WF2 and WF1,
and seem consistent with the observational data quoted above.

For the run without feedback, NoF, over $\sim 75\%$ of the galaxies
are classified as disks. This is because, in the absence of feedback,
the gas cools and flows unimpeded to the center, where it settles into
disks and forms stars profusely. These disks are, however, quite small
(see Fig.~\ref{fig:RhMstr}). The absence of effective feedback allows
the gas to remain undisturbed in such disks, which are quickly
reconstituted after mergers \citep[see, e.g.,][]{Springel2005d,Robertson2006}. At
the other extreme, only $5\%$ of all galaxies in the AGN run, and
$\sim 20\%$ of those in the WF2Dec run, would be classified as disks
according to this criterion.

Strong feedback-driven winds can clearly disturb quiescent disk
morphologies, and their kinematic effects may be difficult to
disentangle from those of ongoing mergers. It remains to be seen
whether a simple feedback model can account for both the observed
frequency of galaxies with disk-like kinematics as well as the
mounting evidence for large scale galactic outflows at $z\sim 2$
\citep[][]{Steidel2010}.

%%%%%%%%%%%%%%%%%%%%%%%%%%
%%%%%%%%%%%%%%%%%%%%%%%%%%

\section{Summary and Conclusions}
\label{sec:conc}

We study the effects of various feedback implementations on the
structure and morphology of simulated galaxies at $z=2$.  Our analysis
uses nine runs from the OverWhelmingly Large Simulations ({\small
OWLS}) project, and probe a variety of possible feedback
implementations, from ``no feedback'' to supernova-driven wind
feedback to strong outflows aided by the contribution from AGNs.
Except for the no-feedback and AGN-feedback cases, all other runs
assume that the {\it same amount} of feedback energy (per mass of
stars formed) is devolved by supernovae to the interstellar medium:
the main difference is {\it how} this energy is coupled to the medium,
which in turn determines the overall effectiveness of the feedback.

Each run follows the evolution of the {\it same} $25 \, h^{-1}$ Mpc
box up to $z=2$, with $512^3$ dark matter particles and $512^3$
particles for the baryonic component. All other simulation parameters
(star formation algorithm, stellar initial mass function, etc) are
kept constant, so any differences between runs may be traced solely to
feedback. In total, we analyze for each run $\sim 150$ galaxies formed
at the centers of haloes with virial mass in the range $10^{11} \,
h^{-1} \, M_\odot< M_{\rm vir} < 3\times 10^{12} h^{-1} \,
M_\odot$. Our main results may be summarized as follows.

\begin{itemize}

\item 
  Varying the feedback implementation can lead to dramatic differences in
  the mass of galaxies formed in a given dark matter halo. The galaxy
  formation efficiency, $\eta_{\rm gal}=M_{\rm gal}/(f_{\rm bar}
  M_{\rm vir})$, varies by roughly an order of magnitude when
  comparing the no-feedback run (NoF, where $\eta_{\rm gal}\sim 0.5$) to
  the AGN+supernova feedback run (AGN, where $\eta_{\rm gal} \sim 0.05$), the
  two extremes probed by our simulations.

\item The ability of feedback to regulate the efficiency of galaxy
  formation in haloes of different mass varies according to the details
  of the adopted numerical implementation of the feedback. Weak or
  ineffective feedback leads to a decrease in galaxy formation
  efficiency with mass, whereas strong feedback curtails
  preferentially the formation of galaxies in low-mass haloes. The
  mass dependence is, however, modest, with variations in $\eta_{\rm gal}$ of less than
  a factor of $\sim 2$ over the (factor of $\sim 30$) mass range
  spanned by haloes in our sample.

\item Feedback results in strong correlations between galaxy mass and
  angular momentum.  This leaves an imprint on galaxy
  morphologies and on the scaling laws relating mass, size, and
  circular velocity.

\item Weak feedback minimizes disturbances to the settling of gas in
  rotationally-supported structures, and favors the formation and
  survival of quiescent {\it gaseous} disks. However, weak feedback
  also allows much of the gas to form stars early in dense
  protogalactic clumps that are later disrupted in mergers as the
  final galaxy assembles. Such mergers also transfer angular momentum
  from the baryons to the halo. The net result is a predominance of
 dense, spheroid-dominated stellar components and a scarcity of
  spatially-extended star-forming disks. 

\item Strong feedback, on the other hand, promotes the formation of
  large, extended galaxies. Indeed, the more efficient the feedback the
  more massive (and therefore, larger) the halo inhabited by a galaxy
  of given stellar mass. It is thus possible to have fairly large
  galaxies of modest stellar mass because, when feedback is strong,
  they inhabit large, massive haloes. The size, mass, and rotation speeds of these extended galaxies
  compare favorably with those reported by the SINS survey. This,
  however, comes at the expense of inhibiting the survival of
  rotationally-supported disks of quiescent kinematics and of
  preventing the formation of compact stellar spheroids.

\item Moderate-feedback runs result in galaxies that follow scaling
  laws that are intermediate between large star-forming disks, such as
  those studied by the SINS collaboration \citep{Forster2009}, and the
  compact, quiescent early-type systems analyzed by
  \citet{vanDokkum2008}. Disk-like morphologies in both gas and stars
  are common in these runs, in numbers that appear commensurate with
  current constraints.

\end{itemize}

Although far from definitive, the results outlined above are encouraging.
Properly calibrated, simple feedback recipes such as the ones we explore
here seem able to produce galaxies with properties in broad agreement with
observation. One should be aware, however, of the numerical sensitivity of
the results to details of feedback implementation. Nevertheless, if
developed in step with observational progress in the characterization of
the high-redshift galaxy population, simulations are likely to become more
and more reliable tools, useful when trying to make sense of the striking
diversity of high-z galaxies in terms of the current paradigm of structure
formation.

%%%%%%%%%%%%%%%%%%%%%%%%%%
%%%%%%%%%%%%%%%%%%%%%%%%%%

\section*{Acknowledgements}
\label{acknowledgements}

LVS thanks the hospitality of the University of Massachusetts and
Kavli Institute for Theoretical Physics, Santa Barbara, where part of
this work was completed.  LVS is grateful to Amina Helmi, Marcel Haas,
Natasha F{\"o}rster Schreiber and Thiago Gon\c{c}alvez for useful
comments and discussions, as well as to Freeke van de Voort for help
with the plotting routine used in Figure 11. LVS also acknowledge
Amina Helmi, NWO and NOVA for financial support.  This research was
also supported in part by the National Science Foundation under Grant
No. PHY05-51164.  The simulations presented here were run on Stella,
the LOFAR BlueGene/L system in Groningen, on the Cosmology Machine at
the Institute for Computational Cosmology in Durham as part of the
Virgo Consortium research programme, and on Darwin in Cambridge.  This
work was sponsored by National Computing Facilities Foundation (NCF)
for the use of supercomputer facilities, with financial support from
the Netherlands Organization for Scientific Research (NWO). This work
was supported by Marie Curie Excellence Grant MEXT-CT-2004-014112 and
by an NWO VIDI grant. We thank useful comments from the anonymous
referee that helped to improve the presentation and clarity of this
paper.

\bibliography{master}

\end{document}